\documentclass[11pt]{article}
\usepackage{url}

\usepackage{hyperref}

\usepackage{amssymb,amscd,amstext,amsmath, amsthm, graphicx}
\usepackage{latexsym}
\usepackage[russian,english]{babel}
\usepackage{floatflt}
\usepackage[mathscr]{eucal}
\usepackage{tikz}
\usetikzlibrary{arrows}
\usepackage{verbatim}
\usepackage{soul}
\usepackage{wrapfig}
\usepackage{fancyhdr}
\usepackage{hyperref}
\usepackage{setspace}
\usepackage{mathdots}
\usepackage{blindtext}
\usepackage{enumitem}
 \usepackage{placeins}
 \usepackage{algorithm}
 \usepackage{csquotes}
\usepackage{multirow}
\usepackage{longtable}
\usepackage[top=1in, bottom=1in, left=1in, right=1in]{geometry}

\usepackage{tikz}\usetikzlibrary{shadows}
\newcommand*\keystroke[1]{%
  \tikz[baseline=(key.base)]
    \node[%
      draw,
      fill=white,
      drop shadow={shadow xshift=0.25ex,shadow yshift=-0.25ex,fill=black,opacity=0.75},
      rectangle,
      rounded corners=2pt,
      inner sep=1pt,
      line width=0.5pt,
      font=\scriptsize\sffamily
    ](key) {#1\strut}
  ;
}

\definecolor{linkcolor}{HTML}{000000}
\definecolor{urlcolor}{HTML}{000000}
\hypersetup{pdfstartview=Fit, linkcolor=linkcolor,urlcolor=urlcolor,
colorlinks=true, citecolor=linkcolor}
\renewenvironment{abstract}{%
\begin{center}\begin{minipage}{0.9\textwidth}\begin{small}
\textbf{Abstract.}}
{\end{small}\par\noindent\end{minipage}\end{center}}

\title{\bf On the Sixth International Olympiad in Cryptography NSUCRYPTO\footnote{The work of the first two authors and the sixth author was supported by Mathematical Center in Akademgorodok under agreement No. 075-15-2019-1613 with the Ministry of Science and Higher Education of the Russian Federation and Laboratory of Cryptography JetBrains Research. The work of the seventh, eighth and eleventh authors was supported by Russian Foundation for Basic Research (projects no. 20-31-70043, 18-07-01394, 19-31-90093).}
}

\author{A.~Gorodilova$^{1}$,
    N.~Tokareva$^{1,2}$,
    S.~Agievich$^{3}$,
    C.~Carlet$^{4}$,
    E.~Gorkunov$^{1,5}$,\\
    V.~Idrisova$^{1}$,
    N.~Kolomeec$^{1}$,
    A.~Kutsenko$^{1,5}$,
    R.~Lebedev$^{5}$,
    S.~Nikova$^{6}$,\\
    A.~Oblaukhov$^{1}$,
    I.~Pankratova$^{7}$,
    M.~Pudovkina$^{8}$,
    V.~Rijmen$^{6}$,
    A.~Udovenko$^{9}$
    \\
  \\
  {\small$^{1}$Sobolev Institute of Mathematics, Novosibirsk, Russia} \\
  {\small$^{2}$Laboratory of Cryptography JetBrains Research} \\
  {\small$^{3}$Belarusian State University, Minsk, Belarus} \\
  {\small$^{4}$University of Paris 8, Paris, France} \\
  {\small$^{5}$Novosibirsk State University, Novosibirsk, Russia}\\
  {\small$^{6}$ESAT-COSIC, KU Leuven, Leuven, Belgium} \\
  {\small$^{7}$Tomsk State University, Tomsk, Russia}\\
  {\small$^{8}$Bauman Moscow State Technical University, Moscow, Russia}\\
  {\small$^{9}$SnT, University of Luxembourg, Esch-sur-Alzette, Luxembourg}\\
    \\
    {\small E-mail: {\tt nsucrypto@nsu.ru}}
    }

\date{}

\begin{document}

\hypersetup{pageanchor=false}

\begin{titlepage}
\maketitle
\begin{abstract}
NSUCRYPTO is the unique cryptographic Olympiad containing scientific mathematical problems for professionals, school and university students from any country. Its aim is to involve young researchers in solving curious and tough scientific problems of modern cryptography. From the very beginning, the concept of the Olympiad was not to focus on solving olympic tasks but on including unsolved research problems at the intersection of mathematics and cryptography. The Olympiad history starts in 2014. In 2019, it was held for the sixth time. In this paper, problems and their solutions of the Sixth International Olympiad in cryptography NSUCRYPTO'2019 are presented. We consider problems related to attacks on ciphers and hash functions, protocols, Boolean functions, Dickson polynomials, prime numbers, rotor machines, etc. We discuss several open problems on mathematical countermeasures to side-channel attacks, APN involutions, S-boxes, etc. The problem of finding a collision for the hash function {\tt Curl27} was partially solved during the Olympiad.

\vspace{0.2cm}

\end{abstract}
\end{titlepage}

\hypersetup{pageanchor=true}
\pagenumbering{arabic}

\section*{Introduction}

NSUCRYPTO (Non-Stop University Crypto) is the International Olympiad in cryptography that was held for the sixth time in 2019.
\begin{floatingfigure}[r]{5.2cm}
\hspace{-0,9cm}
\vspace{-5mm}
\includegraphics[width=0.33\textwidth]{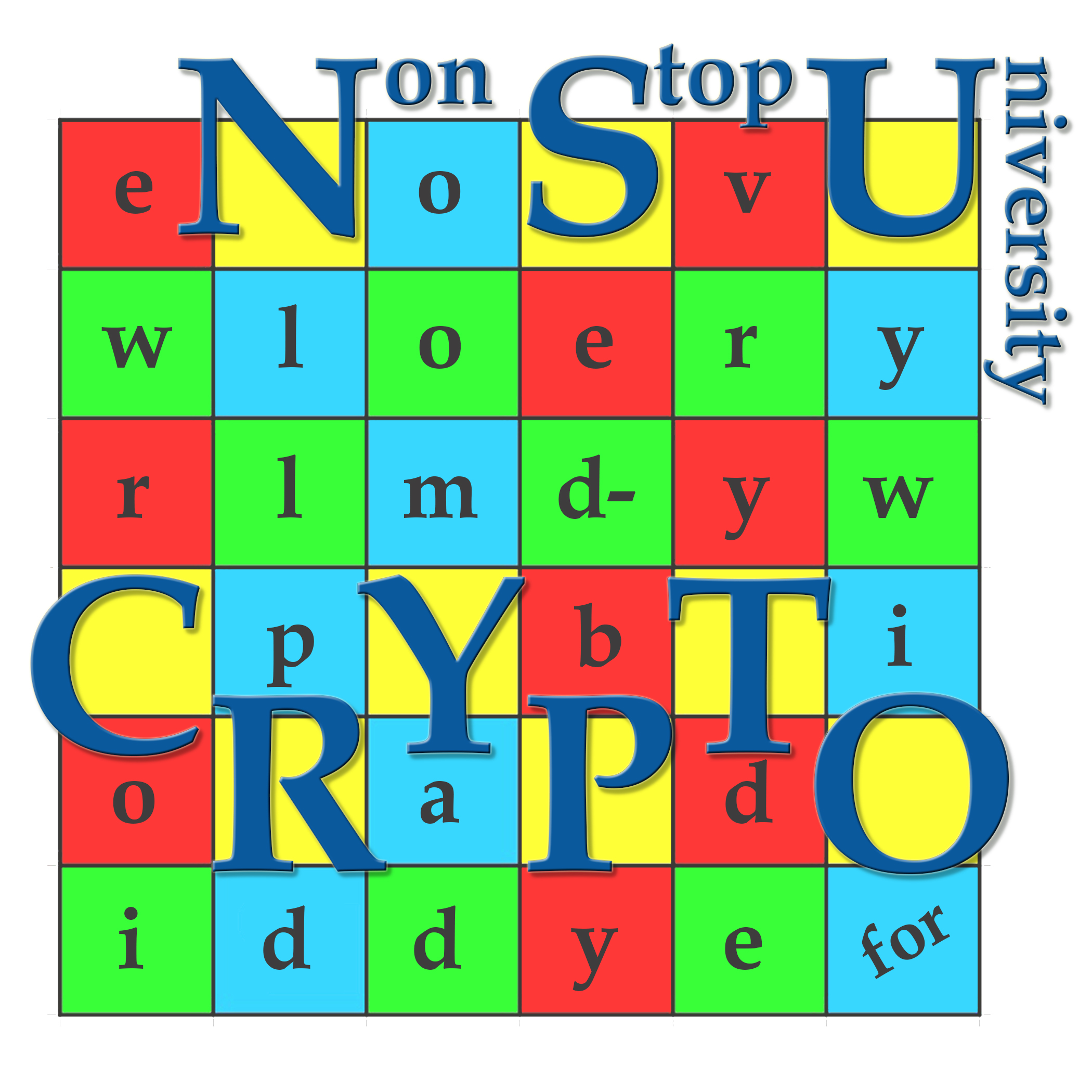}
\label{fig:logo}
\end{floatingfigure}

Interest in the Olympiad around the world is significant. This year, there were hundreds of participants from 26 countries; 42 participants in the first round and 21 teams in the second round from 16 countries were awarded with prizes and honorable diplomas. The Olympiad program committee includes specialists from Belgium, France, the Netherlands, the USA, Norway, India, Luxembourg, Belarus', Kazakhstan, and Russia.

Let us shortly formulate the format of the Olympiad. One of the Olympiad main ideas is that everyone can participate! Each participant chooses his/her category when registering on the Olympiad \href{https://nsucrypto.nsu.ru/}{\textcolor{blue}{website}} \cite{nsucrypto}. There are three categories: ``school students'' (for junior researchers: pupils and high school students), ``university students'' (for participants who are currently studying at universities) and ``professionals'' (for participants who have already completed education or just want to be in the restriction-free category). Awarding of the winners is held in each category separately.

The Olympiad consists of two independent Internet rounds: the first one is individual (duration 4 hours 30 minutes) while the second round is a team one (duration 1 week). The first round is divided into two sections: A --- for ``school students'', B --- for ``university students'' and ``professionals''. The second round is common to all participants. Participants read the Olympiad problems and submit their solutions using the Olympiad website. The language of the Olympiad is English.

The Olympiad participants are always interested in solving different problems of various complexities at the intersection of mathematics and cryptography. They show their knowledge, creativity and professionalism. That is why the Olympiad not only includes interesting tasks with known solutions but also offers unsolved problems in this area. This year, one of such open problems, ``{\tt Curl27}'' (see section~\ref{curl}), was partially solved during the second round! All the open problems stated during the Olympiad history can be found \href{https://nsucrypto.nsu.ru/unsolved-problems}{\textcolor{blue}{here}} \cite{nsucrypto-unsolved}. On the website we also mark the current status of each problem. For example, in addition to ``{\tt Curl27}'', the problem ``Sylvester matrices'' was solved by three teams in 2018, the problem ``Algebraic immunity'' was completely solved during the Olympiad in 2016. And what is important for us, some participants were trying to find solutions after the Olympiad was over. For example, a partial solution for the problem ``A secret sharing'' (2014) was proposed in \cite{sharing}.
We invite everybody who has ideas on how to solve the problems to send your solutions to~us!

The paper is organized as follows. We start with problem structure of the Olympiad in section~\ref{problem-structure}. Then we present formulations of all the problems stated during the Olympiad and give their detailed solutions in section~\ref{problems}. Finally, we publish the lists of NSUCRYPTO'2019 winners in section \ref{winners}.

Mathematical problems and their solutions of the previous International Olympiads in cryptography NSUCRYPTO from 2014 to 2018 can be found in \cite{nsucrypto-2014}, \cite{nsucrypto-2015}, \cite{nsucrypto-2016}, \cite{nsucrypto-2017}, and \cite{nsucrypto-2018} respectively.


\section{Problem structure of the Olympiad}
\label{problem-structure}

There were 16 problems stated during the Olympiad, some of them were included in both rounds (Tables\;\ref{Probl-First},\,\ref{Probl-Second}). Section A of the first round consisted of six problems, whereas the section B contained seven problems. Three problems were common for both sections. The second round was composed of eleven problems. Five problems of the second round included unsolved questions (awarded special prizes from the Program Committee).

\begin{table}[ht]
\centering\footnotesize
\caption{{\bf Problems of the first round}}
\medskip
\label{Probl-First}
\begin{tabular}{cc}
\begin{tabular}{|c|l|c|}
  \hline
  N & Problem title & Maximum scores \\
  \hline
  \hline
  1 & \hyperlink{pr-key}{A 1024-bit key} & 4 \\
    \hline
  2 & \hyperlink{pr-storm}{The magnetic storm}  & 4 \\
    \hline
  3 & \hyperlink{pr-leaves}{Autumn leaves} & 4 \\
    \hline
  4 &  \hyperlink{pr-rotor}{A rotor machine} & 4 \\
    \hline
  5 & \hyperlink{pr-broken-calculator}{Broken {\tt Calculator}} & 4 \\
    \hline
  6 & \hyperlink{pr-promise-sch}{A promise} & 6 \\
  \hline
\end{tabular}
&
\begin{tabular}{|c|l|c|}
  \hline
  N & Problem title & Maximum scores \\
  \hline
    \hline
  1 & \hyperlink{pr-leaves}{Autumn leaves} & 4 \\
    \hline
  2 & \hyperlink{pr-storm}{The magnetic storm} & 4 \\
    \hline
  3 &  \hyperlink{pr-rotor}{A rotor machine} & 4 \\
    \hline
  4 & \hyperlink{pr-16qam}{16QAM} & 8 \\
    \hline
  5 & \hyperlink{pr-promise}{A promise and money}  & 6 \\
    \hline
  6 & \hyperlink{pr-calculator}{{\tt Calculator}} & 6 \\
  \hline
  7 & \hyperlink{pr-apn}{APN + Involutions}  & 7 \\
  \hline
\end{tabular}
\\
\noalign{\smallskip}
Section A
&
Section B
\\
\end{tabular}
\end{table}

\vspace{-0.7cm}

\begin{table}[ht]
\centering\footnotesize
\caption{{\bf Problems of the second round}}
\medskip
\label{Probl-Second}
\begin{tabular}{|c|l|c|}
  \hline
  N & Problem title & Maximum scores \\
  \hline
    \hline
  1 &  \hyperlink{pr-key}{A 1024-bit key} & 4 \\
    \hline
  2 &  \hyperlink{pr-sharing}{Sharing} & 6 + additional scores for open questions\\
    \hline
  3 &  \hyperlink{pr-factoring}{Factoring in 2019}  & 8 \\
    \hline
  4 &  \hyperlink{pr-twinpeaks}{{\tt TwinPeaks-3}}  & 8 \\
    \hline
  5 &  \hyperlink{pr-curl27}{{\tt Curl27}}  & 10 + additional scores for open questions \\
    \hline
  6 & \hyperlink{pr-sbox}{8-bit S-box}   & Unlimited (open problem)  \\
    \hline
  7 &  \hyperlink{pr-rotor}{A rotor machine} & 4 \\
    \hline
  8 &  \hyperlink{pr-16qam}{16QAM}  & 8 \\
    \hline
  9 & \hyperlink{pr-calculator}{{\tt Calculator}}  & 6 \\
    \hline
  10 & \hyperlink{pr-apn}{APN + Involutions (extended)} & 12 + additional scores for open questions \\
    \hline
  11 & \hyperlink{pr-conjecture}{Conjecture}  & Unlimited (open problem) \\
  \hline
\end{tabular}
\end{table}


\section{Problems and their solutions}\label{problems}

In this section, we formulate all the problems of NSUCRYPTO'2019 and present their detailed solutions paying attention to solutions proposed by the participants.

\hypertarget{pr-key}{}
\subsection{Problem ``A 1024-bit key''}

\subsubsection{Formulation}

Alice has a 1024-bit key for a symmetric cipher (the key consists of 0s and 1s). Alice is afraid of malefactors, so she changes her key everyday in the following way:
\begin{itemize}[noitemsep]
\item[{\bf 1.}] Alice chooses a subsequence of key bits such that the first bit and the last bit are equal to 0. She also can choose a subsequence of length 1 that contains only 0.

\item[{\bf 2.}]
Alice inverts all the bits in this subsequence (0 turns into 1 and vice versa); bits outside of this subsequence remain as they are.
\end{itemize}

Prove that the process will stop. Find the key that will be obtained by Alice in the end of the process.

\bigskip

\noindent{\bf Example of an operation.} $11001\underbrace{01101110}011...$ turns to $11001\underbrace{10010001}011...$

\subsubsection{Solution}

Let us encode the binary vector of the key as the corresponding decimal number. It is obvious that this number will increase on the next day, since all the bits on the left from the sequence are not changing, but the first bit of the sequence turns from 0 to 1. Let us note that this number can not increase infinitely since the size of the key is restricted by 1024 bits, so, in the very end the key will be maximal possible and, thus, will consist of all 1s.

Almost all the participants successfully solved the problem.

\begin{figure}[!h]
\centering
\includegraphics[width=0.45\textwidth]{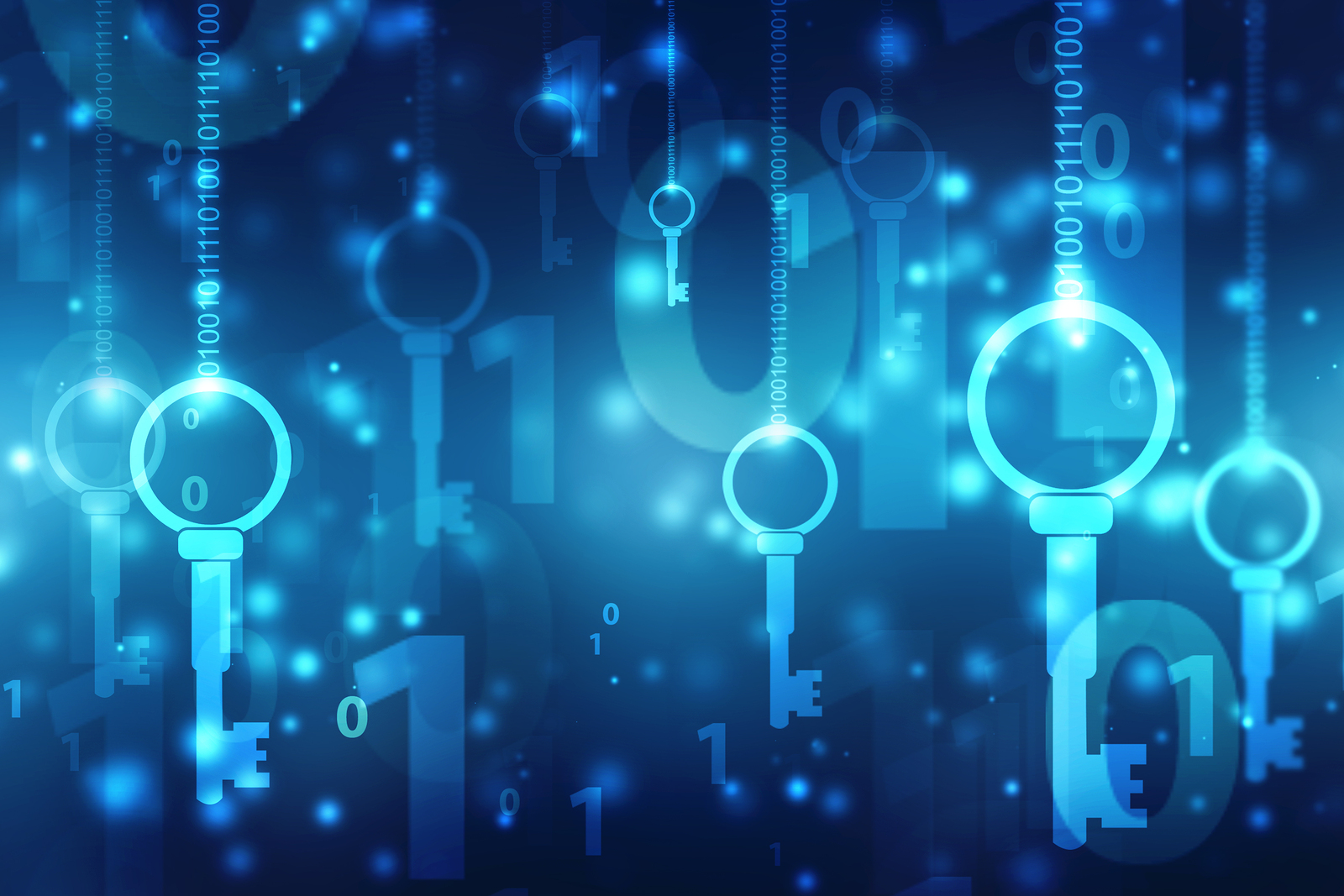}
\end{figure}

\hypertarget{pr-storm}{}
\subsection{Problem ``The magnetic storm''}

\subsubsection{Formulation}

A hardware random number generator is a device that generates random sequences consisting of 0s and 1s. Unfortunately, a disturbance caused by a magnetic storm affected this random number generator. As a result, the device had generated a sequence of 0s of length $k$ (where $k$ is a positive integer), and then started to generate an infinite sequence of 1s.

Prove that at some point the generator will produce a number $1\ldots10\ldots0$ that is divisible by 2019.

\begin{figure}[!h]
\centering
\includegraphics[width=0.45\textwidth]{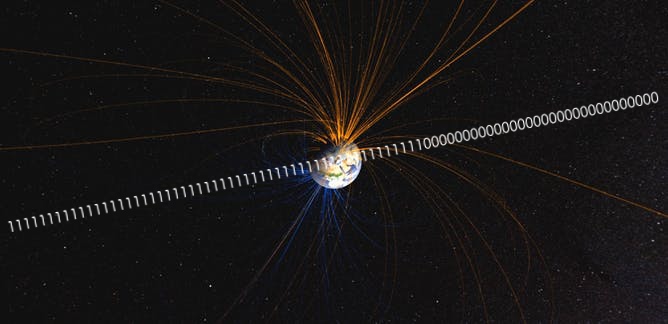}
\end{figure}

\subsubsection{Solution}

Let us prove that a number of form $1\ldots11\ldots1$ is divisible by 2019. Consider all numbers that consists only of 1s, since there are infinite quantity of these numbers, there can be found a pair of numbers A and B such that they have the same remainder when divided by 2019. Therefore, $C=A-B=1\ldots10\ldots0$ consisting of $m$ 1s for some natural $m$ is divisible by 2019, and, since 2019 is not divisible by 2 and 5, $C^*=C\times10\ldots0=1\ldots10\ldots0$ is divisible by 2019 for any number~of~0s.

There were a lot of correct solutions from the participants.

\hypertarget{pr-leaves}{}
\subsection{Problem ``Autumn leaves''}

\subsubsection{Formulation}

Read a hidden message!.. 

\begin{figure}[!h]
\centering
\includegraphics[width=1.0\textwidth]{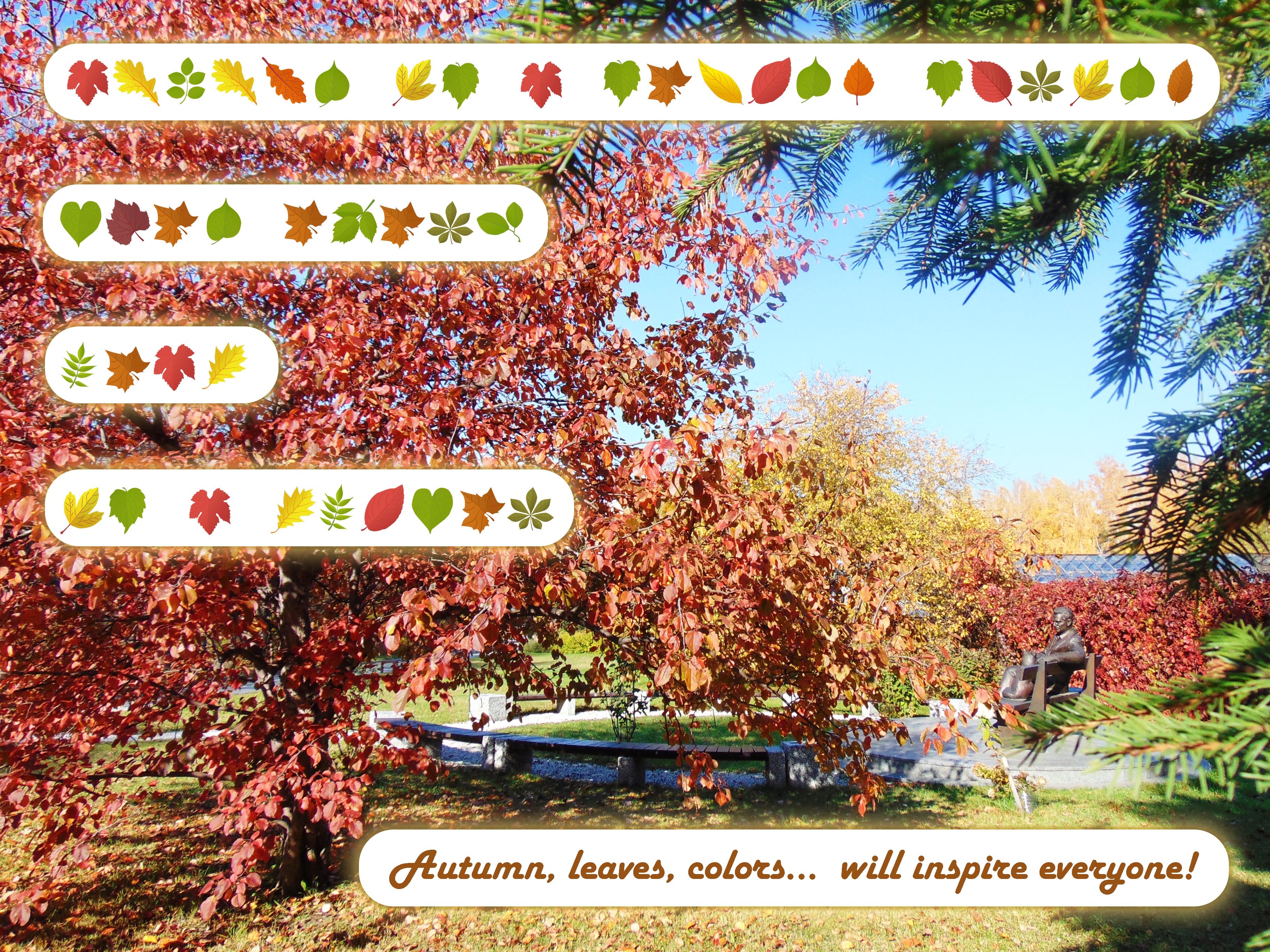}
\label{fig:autumn}
\end{figure}

\subsubsection{Solution}

We see different leaves and spaces between them. It looks like a simple substitution cipher was used there and  distinct leaves corresponded to distinct English letters. By English grammar, we can suppose that the second and the third words are ``{\tt is a}''. Then the first word starts with ``{\tt a}'' and by its structure can be ``{\tt autumn}'' (which is very likely as the autumn landscape is depicted). Also, the leaf~~\includegraphics[width=4mm]{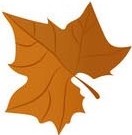}~~is the most common letter in the text and we can guess that it is ``{\tt e}''. Then we see ``{\tt *ea*}'' in the third line that seems to be ``{\tt leaf}''. As a result the last word becomes ``{\tt fl**e*}'' that is ``{\tt flower}''. Finally, we get ``{\tt Autumn is a second spring when every leaf is a flower}'' that is a famous quote by Albert Camus. Almost all the participants read the message.

\hypertarget{pr-rotor}{}
\subsection{Problem ``A rotor machine''}

\subsubsection{Formulation}

In one country rotor machines were very useful for encryption of
information. 

\begin{figure}[!h]
\centering
\includegraphics[width=0.63\textwidth]{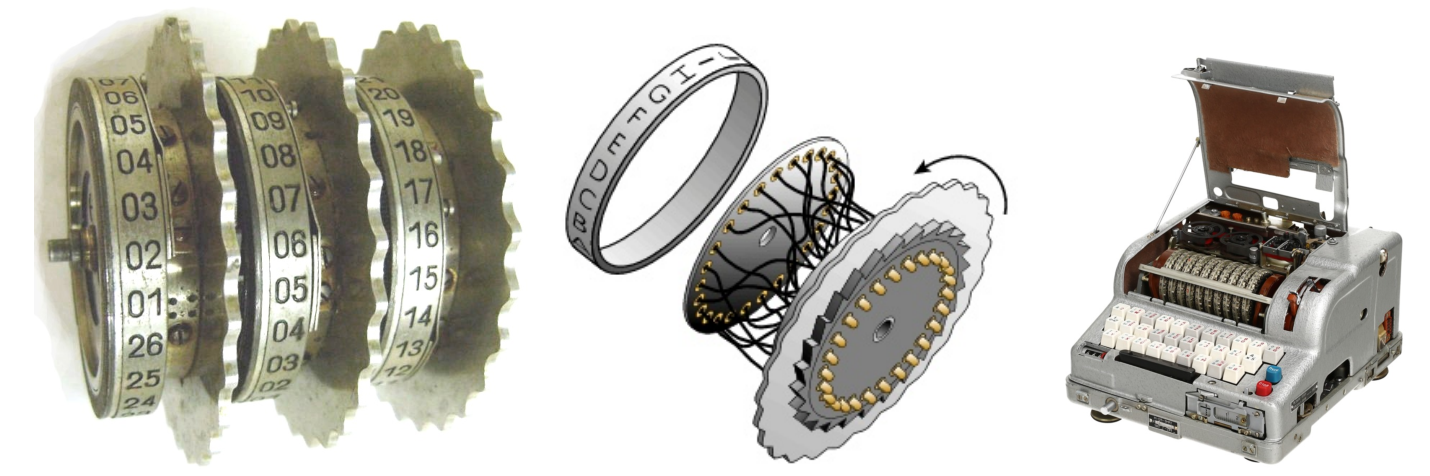}
\end{figure}

Eve knows that for some secret communication a
simple rotor machine was used. It works with letters {\tt O, P, R,
S, T, Y} only and has an input circle with lamps (start), one rotor
and a reflector. See the scheme below. 

\begin{figure}[!h]
\centering
\includegraphics[width=0.63\textwidth]{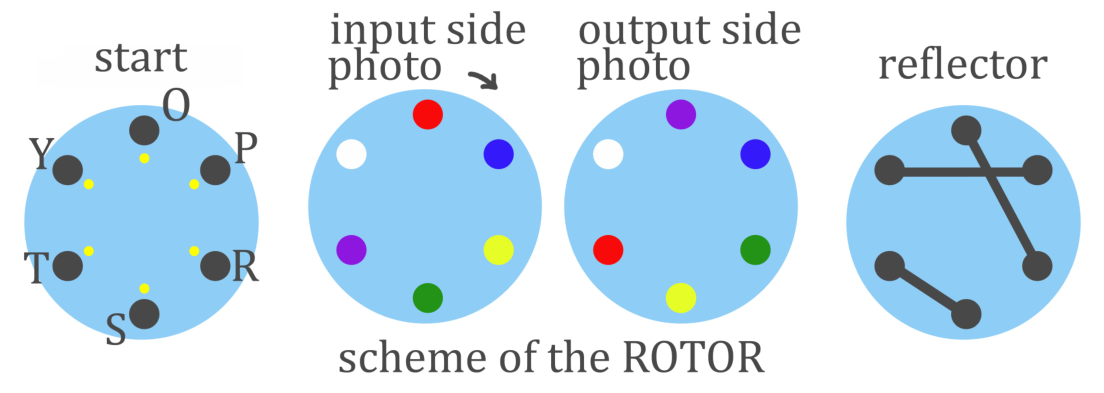}
\end{figure}

The input circle and the reflector are fixed in
their positions while the rotor can be in one of $6$ possible
positions. After pressing a button on a keyboard, an electrical
signal corresponding to the letter goes through the machine, comes
back to the input circle, and the appropriate lamp shows the result
of encryption. After each letter is encrypted, the rotor turns right
(i.\,e. clockwise) on $60$ degrees. Points of different colors on
the rotor sides indicate different noncrossing signal lines within
the rotor.

For instance, if the rotor is fixed as shown on the picture above,
then if you press the button {\tt O}, it will be encrypted as {\tt
T} (the signal enters the rotor via red point, is reflected and then
comes back via purple line). If you press {\tt O} again, it will be
encrypted as {\tt R}. If you press {\tt T} then, you will get {\tt
S} and so on.

Eve intercepted a secret message: {\tt TRRYSSPRYRYROYTOPTOPTSPSPRS}.
Help her to decrypt it keeping in mind that Eve does not know the
initial position of the rotor.

\subsubsection{Solution}

To solve the problem and decrypt the message, one needs to correctly understand the scheme of work. A key for the cipher is the initial position of the rotor. We denote it by a color of the circle on the input side of the rotor that corresponds to the letter {\tt O}.
Table~\ref{Encryption} represents the encryption tables depending on the key.
\begin{table}[ht]
\centering
\caption{{\bf Encryption tables}}
\smallskip
\label{Encryption}
\begin{tabular}{ccc}
\begin{tabular}{c|cccccc}
        & {\tt O} & {\tt P} & {\tt R} & {\tt S} & {\tt T} & {\tt Y}\\
 \hline
 red    & {\tt T} & {\tt Y} & {\tt S} & {\tt R} & {\tt O} & {\tt P}\\
 white  & {\tt R} & {\tt S} & {\tt O} & {\tt P} & {\tt Y} & {\tt T}\\
 purple & {\tt Y} & {\tt R} & {\tt P} & {\tt T} & {\tt S} & {\tt O}\\
\end{tabular}
& ~~~ &
\begin{tabular}{c|cccccc}
        & {\tt O} & {\tt P} & {\tt R} & {\tt S} & {\tt T} & {\tt Y}\\
 \hline
 green  & {\tt S} & {\tt R} & {\tt P} & {\tt O} & {\tt Y} & {\tt T}\\
 yellow & {\tt S} & {\tt T} & {\tt Y} & {\tt O} & {\tt P} & {\tt R}\\
 blue   & {\tt R} & {\tt T} & {\tt O} & {\tt Y} & {\tt P} & {\tt S}\\
\end{tabular}
\end{tabular}
\end{table}

Trying all six possible keys, we find the only one meaningful message {\tt POST TO TOP OOPS SORRY STOP ROTOR} that corresponds to the ``yellow'' key.

Almost all the participants solved the problem. The most interesting solutions were obtained by creating real models for this rotor machine, for example by a school student Varvara Lebedinskaya (The Specialized Educational Scientific Center of Novosibirsk State University), by the team of Kristina Geut, Sergey Titov, and Dmitry Ananichev (Ural State University of Railway Transport).

\hypertarget{pr-broken-calculator}{}
\subsection{Problem ``Broken {\tt Calculator}''}

\subsubsection{Formulation}

Alice and Bob are practicing in developing toy cryptographic applications for smart\-phones. 
This year they have invented {\tt Calculator}  
that allows one to perform the following operations modulo $2019$ (that is to get the result as the reminder of division by 2019):
\vspace{-0.2cm}
\begin{itemize}[noitemsep]
\item to insert at most 4-digit positive integers (digits from 0 to 9);
\item to perform addition, subtraction and multiplication of two numbers;
\item to store temporary results and read them from the memory.
\end{itemize}

\vspace{-0.2cm}
Suppose that Alice wants to send Bob a ciphertext $y$ (given by a 4-digit integer). She sends $y$ from her smartphone to Bob's {\tt Calculator} memory. To decrypt $y$, Bob needs to get the plaintext $x$ (using his {\tt Calculator}) by the rule:
 $x$ is equal to the remainder of dividing ~$f(y)=y^5+1909y^3+401y$~ by 2019.

At the most inopportune moment, Bob dropped his smartphone and broke its screen 
Now, the button
\keystroke{\,+\,} as well as all digits except \keystroke{\ 1\ } and \keystroke{\ 5\ } are not working.

Help Bob to invent an efficient algorithm how to decrypt any ciphertext $y$ using {\tt Calculator} in his situation.
More precisely, suggest a short list of commands, where each command has one of the following types ($1\leqslant j,k<i$):

\vspace{-0.2cm}
\begin{center}
$S_i= y$,~~~~ $S_i=a$,~~~~$S_i=S_j-S_k$,~~~~$S_i=S_j*S_k$,
\end{center}

\vspace{-0.2cm}
\noindent where $a$ is an at most 4-digit integer consisting of digits 1 and 5 only;  for example, $a = 1$, $a = 15$, $a = 551$, $a = 5115$, etc.

The first command has to be $S_1=y$. In the last command, the resulting plaintext~$x$ has to be calculated. We remind that all calculations are modulo $2019$. In particular, the integer~$2500$ becomes~$481$ and $-1000$ becomes~$1019$ immediately after entering or calculations. The shorter the list of commands you suggest, the more scores you get for this problem.

\bigskip

\begin{floatingfigure}[r]{9.2cm}
\hspace{0,5cm}
\includegraphics[width=0.4\textwidth]{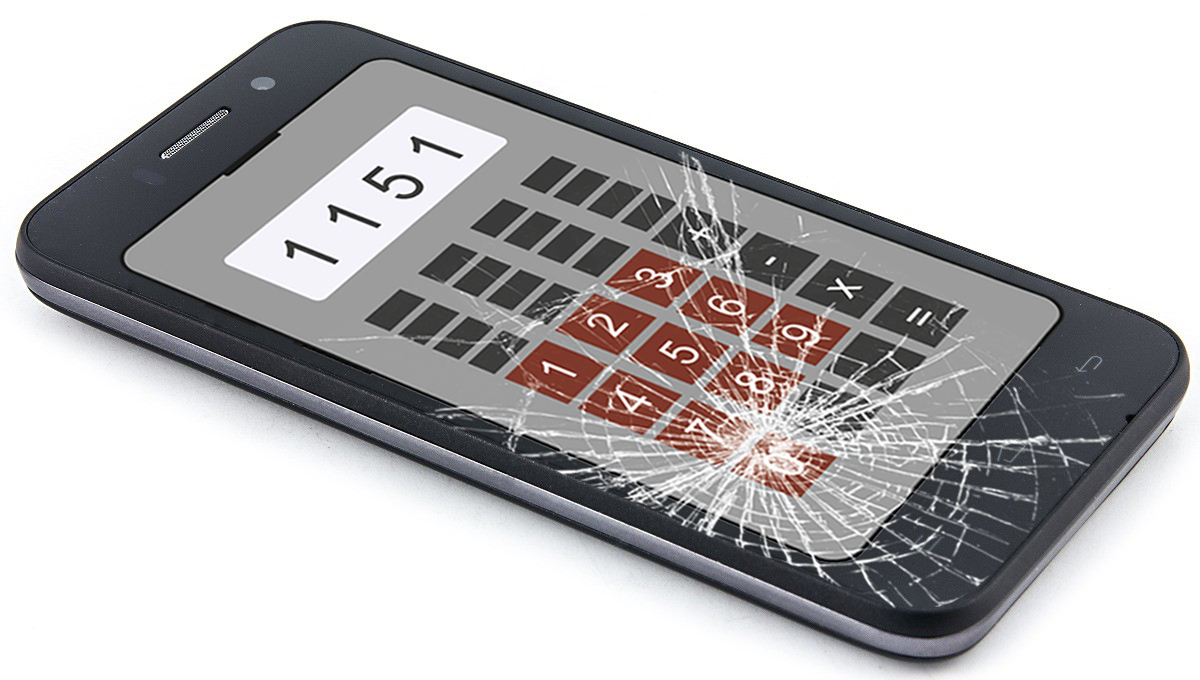}
\label{fig:phone-sch}
\end{floatingfigure}
\noindent {\bf Example.} The following list of commands calculates $x = y^2 - 55$:

\medskip
\begin{tabular}{l|c}
Command & Result\\
\hline
$S_1= y$ & $y$ \\
\hline
$S_2=S_1*S_1$ & $y^2$ \\
\hline
$S_3=11$ & $11$ \\
\hline
$S_4=5$ & $5$ \\
\hline
$S_5=S_3*S_4$ & $55$ \\
\hline
$S_6=S_2-S_5$ & $y^2-55$ \\
\end{tabular}

\subsubsection{Solution}

Let us present the original solution by the programm committee that has 14 steps.

Let $a\equiv_mb$ mean that integers $a$ and $b$ are congruent modulo $m$.
The following relations hold:
\begin{align*}
f(y)&\equiv_{2019}y^5+1909y^3+401y\\
&\equiv_{2019} y( y^4 - 110 y^2 + 401 )\\
&\equiv_{2019} y( y^4 - 2 * 55 y^2 + 55^2 - 55^2 + 401)\\
&\equiv_{2019} y( (y^2 - 55)^2 - 55^2 + 5 * 22^2)\\
&\equiv_{2019} y( (y^2 - 55)^2 - 11^2 * (5^2 - 5 * 2^2))\\
&\equiv_{2019} y( (y^2 - 55)^2 - 11^2 * 5)\\
&\equiv_{2019} y( (y^2 - 55)^2 - 11 * 55).\\
\end{align*}
Thus, the reminder of division of $f(y)$ by 2019 can be calculated for any $y$ by the list of commands given in Table~\ref{Dickson-15}. A similar solution was found by Borislav Kirilov (Bulgaria, The First Private Mathematical Gymnasium).

\begin{table}[!h]
\centering
\caption{{\bf List of commands for the programm committee solution}}
\label{Dickson-15}
\medskip
\begin{tabular}{|l|l||l|l||l|l|}
 \hline
Command & Result & Command & Result & Command & Result \\
 \hline
$S_1= y$           & $y$    & $S_4= S_2 - S_3$ & $y^2-55$     & $S_7= S_3 * S_6$  & $11 * 55$  \\
$S_2= S_1 * S_1$   & $y^2$  & $S_5= S_4 * S_4$ & $(y^2-55)^2$ & $S_8= S_5 - S_7$  & $(y^2 - 55)^2 - 11 * 55$
\\
$S_3= 55$          & $55$   & $S_6= 11$        & $11$         &   $S_{9} = S_1 * S_8$    & $y( (y^2 - 55)^2 - 11 * 55)$\\
 \hline
\end{tabular}
\end{table}

\noindent {\bf Note.} The polynomial $f(y)=y^5+1909y^3+401y$ is the Dickson polynomial $D_5(y,a) = y^5 - 5y^3a + 5 y a^2$ for $a=22$ with coefficients taken modulo~$2019$.

\hypertarget{pr-calculator}{}
\subsection{Problem ``{\tt Calculator}''}

\subsubsection{Formulation}

Alice and Bob are practicing in developing toy cryptographic applications for smart\-phones. This year they have invented {\tt Calculator} that allows one to perform the following operations modulo~$2019$:
\vspace{-0.2cm}
\begin{itemize}[noitemsep]
\item to insert at most 4-digit positive integers (digits from 0 to 9);
\item to perform addition, subtraction and multiplication of two numbers;
\item to store temporary results and read them from the memory.
\end{itemize}

\vspace{-0.2cm}
Suppose that Alice wants to send Bob a ciphertext $y$ (given by a 4-digit integer). She sends $y$ from her smartphone to Bob's {\tt Calculator} memory. To decrypt $y$, Bob needs to get the plaintext $x$ (using his {\tt Calculator}) by the rule $x=f(y)\bmod 2019,$
where $f$ is a secret polynomial known to Alice and Bob only.

At the most inopportune moment, Bob dropped his smartphone and broke its screen 
Now, the button
\keystroke{\,+\,} as well as all digits except \keystroke{\ 2\ } are not working.

Help Bob to invent an efficient algorithm how to decrypt any ciphertext $y$ using {\tt Calculator} in his situation if the current secret polynomial is $f(y)=y^5+1909y^3+401y$.
More precisely, suggest a short list of commands, where each command has one of the following types ($1\leqslant j,k<i$):

\begin{center}
\begin{tabular}{lllllll}
$S_i= y$, &~~~~& $S_i=2$,   &~~~~&  $S_i=222$,   &~~~~&  $S_i=S_j-S_k$,\\
          & &   $S_i=22$,  & &     $S_i=2222$,  & &  $S_i=S_j*S_k$.\\
\end{tabular}
\end{center}

The first command has to be $S_1=y$. In the last command, the resulted plaintext~$x$ has to be calculated. We remind that all calculations are modulo~$2019$. In particular, the integer~$2222$ becomes~$203$ immediately after entering. The shorter the list of commands you suggest, the more scores you get for this problem.

\bigskip

\begin{floatingfigure}[r]{9.2cm}
\hspace{0,5cm}
\includegraphics[width=0.4\textwidth]{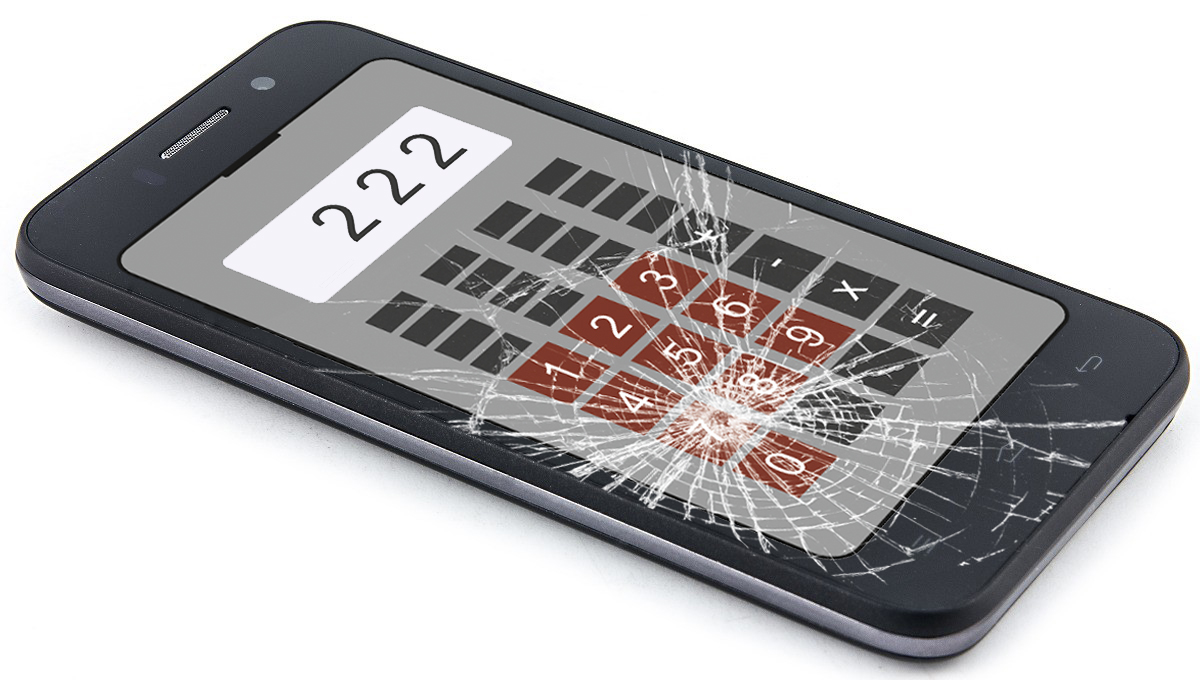}
\vspace{-3mm}
\label{fig:phone}
\end{floatingfigure}
\noindent {\bf Example.} The following list of commands calculates $x = y^2 - 4$:

\medskip
\begin{tabular}{l|c}
Command & Result\\
\hline
$S_1= y$ & $y$ \\
\hline
$S_2=S_1*S_1$ & $y^2$ \\
\hline
$S_3=2$ & $2$ \\
\hline
$S_4=S_3*S_3$ & $4$ \\
\hline
$S_5=S_2-S_4$ & $y^2-4$ \\
\end{tabular}

\subsubsection{Solution}
 The polynomial $f(y)=y^5+1909y^3+401y$ is the Dickson polynomial $D_5(y,a) = y^5 - 5y^3a + 5 y a^2$ for $a=22$ with coefficients taken modulo~$2019$. The following relations hold:
\begin{align*}
D_5(y,a)&=y D_4(y,a)-a D_3(y,a)\\
&=y D_2(D_2(y,a),a^2)-a D_3(y,a)\\
&=y((y^2-2a)^2-2a^2)-ay(y^2-2a-a).
\end{align*}
For $a=22$, the value $f(y)$ can be calculated for any $y$ by the list of commands given in Table~\ref{Dickson}.

\begin{table}[!h]
\centering
\caption{{\bf List of commands for the programm committee solution}}
\label{Dickson}
\medskip
\begin{tabular}{|l|l||l|l|}
 \hline
Command & Result & Command & Result \\
 \hline
$S_1= y$               & $y$        &   ~$S_8    = S_7 * S_7$       & $(y^2-2a)^2$\\
$S_2= 2$               & $2$        &   ~$S_{9} = S_8 - S_5$       & $(y^2-2a)^2-2a^2$\\
$S_3= 22$              & $a$        &   $S_{10} = S_1 * S_{9}$    & $y((y^2-2a)^2-2a^2)$\\
$S_4= S_2 * S_3$       & $2a$       &   $S_{11} = S_7 - S_2$       & $y^2-2a-a$\\
$S_5= S_3 * S_4$       & $2a^2$     &   $S_{12} = S_1 * S_{11}$    & $y(y^2-2a-a)$\\
$S_6= S_1 * S_1$       & $y^2$      &   $S_{13} = S_3 * S_{12}$    & $ay(y^2-2a-a)$\\
$S_7= S_6 - S_4$       & $y^2-2a$   &   $S_{14} = S_{10} - S_{13}$ & $f(y)$\\
 \hline
\end{tabular}
\end{table}

What was surprising that the participants found two solutions that has 11 and 13 steps! These solutions were awarded by additional points.
The solution with 11 steps were found by Madalina Bolboceanu (Romania, Bitdefender) during the first round (Table~\ref{11-steps}). The solution with 13 steps were given by Henning Seidler and Katja Stumpp team (Germany, TU Berlin) during the second round. Both of the solution were based on the representation $f(y) = y((y^2-44)(y^2-66) - 22^2)$.

\begin{table}[!h]
\centering
\caption{{\bf List of commands for the 11-step solution}}
\label{11-steps}
\medskip
\begin{tabular}{|l|l||l|l|}
 \hline
Command & Result & Command & Result \\
 \hline
$S_1= y$               & $y$        &   ~$S_7   = S_6 - S_4$       & $y^2-44-22$\\
$S_2= S_1 * S_1$       & $y^2$      &   ~$S_{8} = S_6 * S_7$       & $(y^2-44)*(y^2-44-22)$\\
$S_3= 2$               & $2$        &   $S_{9} = S_4 * S_4$        & $22^2$\\
$S_4= 22$              & $22$       &   $S_{10} = S_8 - S_9$       & $(y^2-44)*(y^2-44-22) - 22^2$\\
$S_5= S_3 * S_4$       & $44$       &   $S_{11} = S_1 * S_{10}$    & $f(y)$\\
$S_6= S_2 - S_5$       & $y^2-44$   &       & \\

 \hline
\end{tabular}
\end{table}

\hypertarget{pr-promise-sch}{}
\subsection{Problem ``A promise''}

\subsubsection{Formulation}

Young cryptographers, Alice, Bob and Carol, are interested in quantum computings and really want to buy a quantum computer.
A millionaire gave them a certain amount of money (say, $X_A$ for Alice, $X_B$ for Bob, and $X_C$ for Carol). He also made them promise that they would not tell anyone, including each other, how much money everyone of them had received.

\begin{itemize}[noitemsep]
\item
Could you help the cryptographers to invent an algorithm how to find out (without breaking the promise) whether the total amount of money they have, $X_A+X_B+X_C$, is enough to buy a quantum computer?
\item
What weaknesses does your algorithm have (if someone breaks the promise)? Does it always protect the secret of the honest participants from the dishonest ones?
\end{itemize}

\subsubsection{Solution}

This problem is a particular case for the problem ``A promise and money'' for only three participants (see section~\ref{promise}).

\hypertarget{pr-promise}{}
\subsection{Problem ``A promise and money''}\label{promise}

\subsubsection{Formulation}

A group of young cryptographers are interested in quantum computings and really want to buy a quantum computer.
A millionaire gave them a certain amount of money (say, $n$ cryptographers; $X_i$ for each of them, $i=1,\ldots,n$). He also made a promise from them that they would not tell anyone, including each other, how much money everyone of them had received.

\begin{itemize}[noitemsep]
\item
Could you help the cryptographers to invent an algorithm how to find out (without breaking the promise) whether the total amount of money they have, $\sum_{i=1}^nX_i$, is enough to buy a quantum computer?
\item
What do you think whether there are such algorithms protecting the secrets of honest participants from dishonest ones?
\item
What weaknesses does your algorithm have (if someone breaks the promise)? Does it always protect the secret of honest participants from dishonest ones?
\end{itemize}

\subsubsection{Solution}

Here we give an idea of the solution proposed by Mikhail Kudinov (Bauman Moscow State Technical University).

\begin{floatingfigure}[r]{4.5cm}
\hspace{-0,7cm}
\centering
\includegraphics[width=0.25\textwidth]{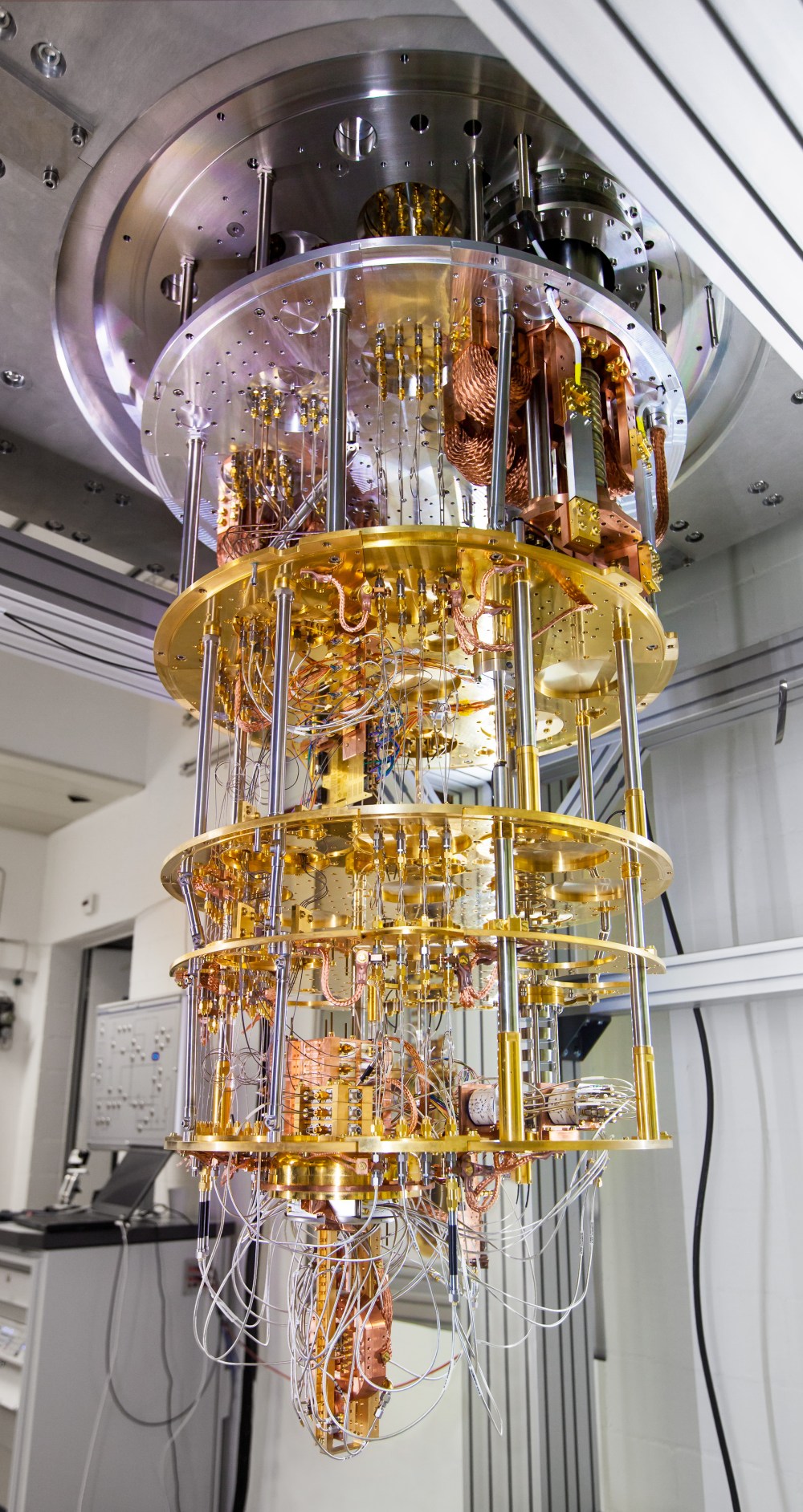}
{\footnotesize {\bf IBM's 50 qubit quantum computing system \cite{IBM}}}
\label{fig:qc}
\end{floatingfigure}

First of all, it is supposed that no one can buy a quantum computer himself without other participants.
Let us assume that $N'$ is the amount of money that one needs to buy a quantum
computer and $N = n N'$, where $n$ is the number of participants. The millionaire gave them $X_i$ money for $i \in \{1, \ldots, n\}.$
Each of participants chooses random secrets $s_{i,j}$ uniformly so that
$$
	\sum _{j=1}^{n} s_{i,j} \equiv X_i \pmod{N}.
$$
Then each of then gives the share $s_{i,j}$ to the owner of $X_j$ by the
secure channel. After this procedure, the owner of $X_i$ has shares $s_{k,i}$ for each $k  \in \{1, \ldots, n\}$.
It is obvious that
$$
	\sum_{j=1}^{n}\sum_{i=1}^{n} s_{i,j} = \sum_{i=1}^{n} X_i \pmod{N}.
$$
Under the first suggestion, all participants can together calculate the common amount of money.

The main disadvantage of the algorithm, in addition to the suggestion, is a big amount of private communication (though the number of keys can be $n$ for asymmetric schemes).

Analogically, many participants described algorithms similar to Schneier's calculating average salary algorithm~\cite{Schneier}.
In general, all such algorithms are vulnerable if $n-1$ participants are dishonest.
Some participants tried to describe a possibility to use a cryptosystem, that is homomorphic by~``$+$'' and preserves relation~``$<$'', as a general analysis.

The problem of the first school round is the same problem for $n = 3$ (score assignment was more loyal).
Despite there was a quite big number of solutions for this problem in the student round, each solution had big or small lacks in analysis of the general case, in analysis of the algorithm advantages and disadvantages, in description of communications (number of private communications, what kind of cryptography is used, number of required private keys) and so on. As a result, there was no possibility to chose ``best of the best'' for $6$ scores and we decided to give $5$ scores as maximum. There were nine maximal-scored solutions.

\hypertarget{pr-16qam}{}
\subsection{Problem ``16QAM''}

\subsubsection{Formulation}

For sending messages, Alice and Bob use a fiber-optic communication via 16QAM technology.
This technology allows to send messages whose alphabet consists of 16 letters, where each letter is usually encoded with a 4-bit Gray code.
While a message is transmitted in the channel, single errors in codewords of the Gray code are possible.

Alice has read an interesting book and would like to share her enthusiasm with Bob!
Alice sent a short fragment from the book to Bob.
Due to the characteristics of the communication channel used,
she divided the text into two parts and sent them separately.
In the first part, she placed all of the 16 consonants that occurred in this fragment;
in the second part, she placed vowels (``y'' is a vowel), a space, a hyphen and punctuation marks.
Then Alice also encoded the letters with Hamming code to be able to correct single errors.
She applied a 7-bit Hamming code with the parity-check matrix whose columns are written
in lexicographical order.

Bob received the following two parts of ciphertext (given in hexadecimal notation):

\begin{center}
\begin{tabular}{p{7.5cm}p{7.5cm}}
Part 1 & Part 2\\

{\tt
66674C36666F43D3C199900AA1AA325992A\newline
67A59D9B4A8B69330D1BC000153367A5E33\newline
D30E6692D0F349D3321FFFF0ED706667A7F\newline
670D999679F4AA67561BA679B4AA54F34D5\newline
AB0F4AACCF000055CE633670D9DA54CE37F\newline
660DE19CD995335495523CCAAA8F1E03325\newline
86CF48A98CD9B387FD9D546A99E9D200033\newline
3201513FE5B4AA00CCCE9667554CD2CCCB3\newline
330F32A666553CD756AC3E0674E9D369E1D\newline
C6A9999780007F00961E66465519FEA8B25\newline
14CCCB332AA63332CCCE6D2A99AACCCC004
}

&

{\tt
66CA61967319CCD2CE76998CE6433332D19\newline
B46784C65334E999A402ADA0265A99A6633\newline
33319B32D3299698CCC96986619967134CC\newline
B4CE23333334CC6730CE90170CCCD2CE669\newline
996A61999EA63332CCA4C3332D4CD3334CC\newline
D3319994730CCCD3A6669D96A66999699B3\newline
98640CC86CE619676AD4CD3308999866D33\newline
79321C33210B4C6732199B53218019A404C\newline
D2DE65A986663398CCCCCB5319CC6665997\newline
B96A63398CD9CCD2CD9A399A66339866619\newline
98CD9CC325A6339CCE619998C04C66CE633\newline
996A61998CF66967334CC66CA6199865E$(0)_2$
}
\\
\end{tabular}
\end{center}

Also, he received the following number sequence: 22, 19, 3, 3, 36, 53, 3, 33, 20, 28. Each number indicates how many consonants are contained between the punctuation marks.

Recover the text and find the main character of the book Alice has read!

\begin{figure}[!h]
\centering
\includegraphics[width=0.5\textwidth]{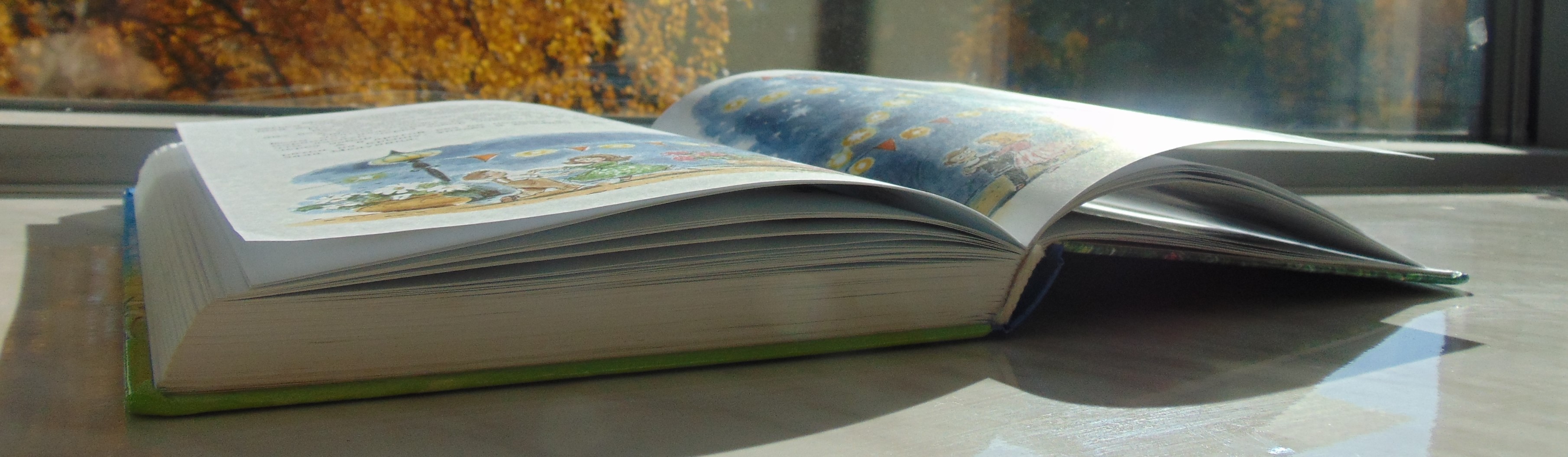}
\vspace{-3mm}
\label{fig:book}
\end{figure}

\subsubsection{Solution}


Some details in the problem statement are insignificant.
Namely, we could omit the step with the Gray code and mind that Alice substitutes
$7$-bit codewords of the Hamming code for each symbol in each part of the plaintext.

The crucial idea to broke the cipher Alice and Bob use is analyzing the frequency distribution in each part of the ciphertext.
This helps to deduce the probable meaning of the most common symbols and form partial words.
Tentative search for combinations of consonants and vowels giving actual words in English expands the partial solution.
Frequencies of pairs of letters also give an improvement but it could seem inessential.
At last, one can employ search engine on the Internet to find the fragment of the book that Alice sent to Bob.

Let us consider a possible solution. Alice uses the Hamming code with the parity check matrix $H$ and the corresponding generator matrix $G$, where
$$
H=\begin{bmatrix}
	0&0&0&1&1&1&1\\
	0&1&1&0&0&1&1\\
	1&0&1&0&1&0&1
\end{bmatrix},
~~
G=\begin{bmatrix}
	1&1&1&0&0&0&0\\
	1&0&0&1&1&0&0\\
	0&1&0&1&0&1&0\\
	1&1&0&1&0&0&1
\end{bmatrix}.
$$
First, rewrite each part of the given ciphertext in the binary form.
Split them into $7$-bit words and correct errors using the parity check matrix $H$.
One can decode the Hamming code into a $4$-bit Gray code but it is not a necessary step for the solution.
Calculating frequencies of codewords separately in each part of the given ciphertext, we put them in Table~\ref{Freq}.

\begin{table}[ht]
\centering\footnotesize
\caption{{\bf Frequencies of Hamming codewords in the text}}
\smallskip
\label{Freq}
\begin{tabular}{cc}
\begin{tabular}{|c|c|c|}
		\hline
		Gray code & Hamming code & Frequency\\
		\hline
		1011 & 0110011 & 46\\ 0010 & 0101010 & 30\\ 1001 & 0011001 & 24\\ 0001 & 1101001 & 24\\
		\hline
		0011 & 1000011 & 19\\ 0000 & 0000000 & 15\\ 0110 & 1100110 & 13\\ 1100 & 0111100 & 8\\
		\hline
		1111 & 1111111 & 8\\ 1101 & 1010101 & 7\\ 0100 & 1001100 & 6\\ 1110 & 0010110 & 5\\
		\hline
		1010 & 1011010 & 5\\ 0101 & 0100101 & 4\\ 1000 & 1110000 & 4\\ 0111 & 0001111 & 2\\
		\hline
\end{tabular}
&
\begin{tabular}{|c|c|c|}
\hline
		Gray code & Hamming code & Frequency\\
		\hline
		0100 & 1001100 & 85\\ 1011 & 0110011 & 50\\ 1001 & 0011001 & 33\\ 0001 & 1101001 & 26\\
		\hline
		1010 & 1011010 & 17\\ 0011 & 1000011 & 9\\0000 & 0000000 & 8\\ 1110 & 0010110 & 7\\
		\hline
		1100 & 0111100 & 2\\ 0010 & 0101010 & 1\\ 1000 & 1110000 & 1\\ 0111 & 0001111 & 0\\
		\hline
		0101 & 0100101 & 0\\ 1101 & 1010101 & 0\\ 0110 & 1100110 & 0\\ 1111 & 1111111 & 0\\
		\hline
\end{tabular}
\smallskip
\\

a) Part 1 & b) Part 2
\end{tabular}
\end{table}

Compare the frequencies obtained with those of letters in the English language.
The suitable frequency distribution can be found in~\cite{Lewand2000}, which is cited, e.~g., at~\cite{frequency}.
According to Lewand, arranged from most to least common in appearance, the letters are:
$$\texttt{e t a o i n s h r d l c u m w f g y p b v k j x q z}.$$

We start with vowels, punctuation marks, spaces, and a hyphen, which are placed in Part~2.
Make a guess that the most frequent symbol in Part~2 is the space.
It is also worth to note that most of punctuation marks are followed by a space in contrast to a hyphen,
which is usually embraced by letters.
Using letter frequencies, we determine the probable spaces, vowels, and hyphen,
and construct the following partial solution for this part of the plaintext (the sign \verb|#| substitutes punctuation):

\smallskip
{\small
{\tt ee ae e oe o e ua iaia\# e oo oy-oy i o ea ee\# u\# ea\# auae o ie ea o e aoy a oe \par
 o i a i eae\# a i o o o eae a oo o i o iee ay ue aeii o aa aie\# uuay\# e uai uy oy \par
 oe i a e ea i e eae\# i e ee oeee o e a a ee a\# e e a uy ee e i a e oe o ee a a\#}
}

\smallskip
Let us turn to Part~1, which contains $16$ consonants occurring in the fragment of the book.
Let us order the codewords of the Hamming code from most to least frequent in Part~1, as it is shown in Table~\ref{Freq}a.
Denote the $7$-bit codewords by hexadecimal numbers from \texttt{0} till \texttt{F}.
Then we get the following ciphertext of $220$ symbols in length that is splitted into $10$ pieces (according to the number sequence given in the task):

\smallskip
{\footnotesize
\verb|023402C43E0251412B0103| ~~
\verb|02C1B32407551003703| ~~
\verb|4A3| ~~
\verb|B46| ~~
\verb|33A4884CE02E804020631094106311739943|\par
\verb|1675510A0040C1068047266101D10619FF56D4031A00048090103| ~~
\verb|355|\par
\verb|025108B315023021A3020246102173994| ~~
\verb|E2333C72410275585D46| ~~
\verb|021281BD102021A0202631016055|\par
}

\smallskip
Then we match symbol frequencies in Part~1 of the ciphertext with those of consonants in the English alphabet.
The first five pairs are like as follows:
{\tt 0 - t}, {\tt 1 - n}, {\tt 2 - s/h}, {\tt 3 - s/h}, {\tt 4 - r}.

\smallskip
The bigram \texttt{th} is the most frequent in English. This allows us to make a suggestion that \texttt{2} substitutes \texttt{h} and \texttt{3} substitutes \texttt{s}.
Then we obtain a partial solution for Part~1 and, combining with one for Part~2, get the following pieces of the plaintext given in Table~\ref{parts}.
It is not difficult to recognize words \texttt{these are the} at the beginning in~(1).
Also, we can see \texttt{the} as the first word in~(2) and~(8).
\vspace{-6pt}
\begin{table}[ht]
\centering\small
\caption{{\bf Partial plaintext}}\label{parts}
\smallskip
\begin{tabular}{|c|p{10.6cm}|}
	\hline
	No. & Partial plaintext \tabularnewline
	\hline
	(1) & \verb|thsrthCrsEth5nrnhBtnts|\newline \verb|ee ae e oe o e ua iaia# | \tabularnewline
	\hline
	(2) & \verb|thCnBshrt755ntts7ts|\newline \verb|e oo oy-oy i o ea ee# | \tabularnewline
	\hline
	(3) & \verb|rAs|\newline \verb|u# | \tabularnewline
	\hline
	(4) & \verb|Br6|\newline \verb|ea# | \tabularnewline
	\hline
	(5) & \verb|ssAr88rCEthE8trtht6snt9rnt6snn7s99rs|\newline \verb|auae o ie ea o e aoy a oe o i a i eae# | \tabularnewline
	\hline
	(6) & \verb|n6755ntAttrtCnt68tr7h66ntnDnt6n9FF56DrtsnAtttr8t9tnts|\newline \verb|a i o o o eae a oo o i o iee ay ue aeii o aa aie# | \tabularnewline
	\hline
	(7) & \verb|s55|\newline \verb|uuay# | \tabularnewline
	\hline
	(8) & \verb|th5nt8Bsn5thsthnAsththr6nthn7s99r|\newline \verb|e uai uy oy oe i a e ea i e eae# | \tabularnewline
	\hline
	(9) & \verb|EhsssC7hrnth75585Dr6|\newline \verb|i e ee oeee o e a a ee a# | \tabularnewline
	\hline
	(10) & \verb|thnh8nBDnththnAthth6sntn6t55|\newline \verb|e e a uy ee e i a e oe o ee a a#| \tabularnewline
	\hline
\end{tabular}
\end{table}

The best idea for the next step is to search through the English dictionary for words that have given vowels in the prescribed order.
It is possible to use one of the tools for pattern recognition available on the Internet, e.~g.,~\cite{pattern}.
Advanced participants of the Olympiad implemented some computer programs on their own.

Consider several examples. We have a word with consonants \texttt{s55} and vowels \texttt{uuay} in~(7),
and the last two consonants are identical. The only match is \texttt{usually}, so we assume that \texttt{5} substitutes the letter \texttt{l}.
The pattern \texttt{auae} in combination with double \texttt{s} gives us two possibilities in~(5)~-- \texttt{assuage} and \texttt{sausage}.
In any case, it seems like \texttt{A} means \texttt{g}. Then we have \texttt{rugs} in~(3).
The pattern \texttt{uai} and consonants \texttt{5nt8B} lead us to \texttt{lunatic} in~(8), so~\texttt{8} probably means \texttt{c}.

At this point we revise our matching the letters and their frequencies corresponding to the Part~1 of the ciphertext.
Let us look at the first eight letters with large frequencies: \texttt{t n h s r l 6 7/c}.
We can see that the letter \texttt{d} has still been hidden.
According to the Lewand distribution it~is the~most probable that \texttt{6} means \texttt{d}.
Then~(4) contains \texttt{Brd} and \texttt{ea} what gives us possible words \texttt{beard} and \texttt{bread}.
Therefore, it seems like \texttt{B} substitutes \texttt{b}.

A thoroughly analysis of the remaining ciphertext and search for words by patterns and number of letters eventually lead us to the plaintext (with punctuation replaced by \verb|#|):

\smallskip
{\tt these are the mores of the lunar inhabitants\#
the moon boy-shorty will not eat \par
sweets\# rugs\# bread\# sausage or ice cream of the factory that does not print\par
 ads in newspapers\#
and will not go to treatment a doctor who did not invented \par
any puzzle advertising to attract patients\#
usually\#
the lunatic buys only \par
those things that he read in the newspaper\#
if he sees somewhere on the wall \par a clever ad\#
then he can buy even the thing that he does not need at all\#}

\smallskip
This is a fragment of the fairytail novel ``Dunno on the Moon'' by Russian writer Nikolay Nosov.
The title character of the novel is a boy-shorty Dunno.
The problem was completely solved by 13 teams in the second round and by Samuel Tang (Hong Kong,	 Black Bauhinia) in the first round. The best solutions were proposed by the team of Irina Slonkina, Mikhail Sorokin, and Vladimir Bobrov (Bauman Moscow State Technical University), and the team of Vladimir Paprotski, Dmitry Zarembo, and Karina Kruglik (Belarusian State University).
\otherlanguage{english}{~}

\hypertarget{pr-apn}{}
\subsection{Problem ``APN + Involutions''}

The first three questions {\bf Q1, Q2, Q3} were given as the problem ``APN + Involutions'' in the first round. The extended version of the task for the second round included also the question {\bf Q4} that contains open problems.

\subsubsection{Formulation}

Alice wants to construct a block cipher with heavy use of {\bf involutions} as subcomponents; this minimizes difference between the algorithms for encryption and decryption. She knows that {\bf APN permutations} are the best choice of subcomponents to resist attacks based on differential technique. She wants to construct a set of APN permutations that are involutions for every $n \geqslant 2$.

Alice knows that any involution can be expressed as the product of disjoint {\bf transpositions}. So, she decides to study the following involution
$$g = \prod\limits_{i = 1}^d {\left( {{{\alpha}_i},{{\alpha '}_i}} \right)}, $$
where $\left\{ {{\alpha _i},{{\alpha '}_i}} \right\} \cap \left\{ {{\alpha _j},{{\alpha '}_j}} \right\} = \emptyset $ for all $i,j \in \{ 1,...,d\} $, $i \ne j$, $1 \leqslant d \leqslant {2^{n - 1}}$.

\medskip

Alice needs your help to get APN permutations among such involutions $g$. Find answers to the following questions!

\begin{enumerate}
  \item[{\bf Q1}] {
  Let
$$\Lambda (g) = \big\{ {{\alpha _i} \oplus {{\alpha '}_i}:\ i = 1,...,d } \big\},
~~~
\widehat \Lambda (g) = \big[ {\alpha _i} \oplus {{\alpha '}_i}:\ i = 1,...,d \big],
$$
$$
{\text{B}}(g) = \big\{ {x  \oplus y:\ \{ x,y\}  \subseteq {\rm FixP}(g)},\ x \neq y  \big\},
~~~
\widehat {{\rm B}}(g) = \big[  {x  \oplus y:\ \{ x ,y\}  \subseteq {\rm FixP}(g)},\ x \neq y \big],$$

where ${\rm FixP}(g)$ is the set of all {\bf fixed points} of $g$, i.\,e.
  ${\rm FixP}(g) = \left\{ {x  \in {\mathbb{F}}_2^n:\ g(x) = x } \right\}.$

Suppose that $g$ is an APN permutation. Get necessary conditions for multisets $\widehat \Lambda (g)$, $\widehat {{\rm B}}(g)$ and sets $\Lambda (g)$, ${\rm B}(g)$. Prove that if your conditions do not hold, then $g$ is not an APN~permutation.
    }

\item[{\bf Q2}] {
  Let
$${{\rm d}_{a,b}}(g) = |\{x  \in {\mathbb{F}}_2^n: g(x  \oplus a) \oplus g(x ) = b \}|,\ \    a,b  \in {\mathbb{F}}_2^n.$$
Let $g$ be an involution and APN.
Find ${{\rm d}_{a ,a }}(g)$ for each nonzero $a\in {\mathbb{F}}_2^n$.

  }
\item[{\bf Q3}]
Can you get the nontrivial upper bound on $| {{\rm FixP}(g)}|$?

\item[{\bf Q4}]{

 Let ${M_n}$ be the set of all $n$-bit involutions that are APN permutations.
  \begin{enumerate}
    \item {Can you find the cardinality of ${M_n}$ for $n = 2,3,4$?}
    \item {Can you find the cardinality of ${M_n}$ for $n = 5$?}
    \item {{\bf Bonus problem (extra scores, a special prize!)}

    Let $n \geqslant 6$. Can you get the lower and the upper bounds for the cardinality of ${M_n}$? Can you describe involutions from ${M_n}$? Can you suggest constructions for involutions from~${M_n}$?  }
  \end{enumerate}

   }
   Note that the mapping $x \mapsto {x^{ - 1}}$ in the Galois field $GF({2^n})$ belongs to ${M_n}$ for odd $n \geqslant 3$.
\end{enumerate}

\noindent{\bf Remark.} Let us recall relevant definitions.
\begin{itemize}[noitemsep]
\item ${\mathbb{F}}_2^n$ is the vector space of dimension over ${{\mathbb{F}}_2} = \{ 0,1\} $.
\item A vector $x \in {\mathbb{F}}_2^n$ has the form $x = ({x_1},...,{x_n})$, where ${x_i} \in {{\mathbb{F}}_2}$.
For two vectors $x,y \in {\mathbb{F}}_2^n$ their sum is $x \oplus y = ({x_1} \oplus {y_1},...,{x_n} \oplus {y_n})$, where $ \oplus $ stands for XOR operation.

\item Let $\widehat X = \big[ {{x_1},...,{x_d}} \big]$ be a multiset with the underlying set ${\mathbb{F}}_2^n$, where ${x_1},...,{x_d}\in{\mathbb{F}}_2^n$.

Note that all elements in a set are distinct. Unlike a set, a multiset allows for multiple instances for each of its elements.
\item A {\bf permutation} $s$ is a mapping from $\mathbb{F}_2^n$ to $\mathbb{F}_2^n$ such that $s(x)\neq s(y)$ for all  $x,y\in\mathbb{F}_2^n$, $x\neq y$.
\item An {\bf involution} $s$ is a permutation that is its own inverse, ${s^2}(x) = s(s(x)) = x$ for all $x \in \mathbb{F}_2^n$.
\item For any different vectors $\alpha ,\beta  \in {\mathbb{F}}_2^n$, a permutation $s$ is called a {\bf transposition} if $s(\alpha ) = \beta $, $s(\beta ) = \alpha $ and $s(x) = x$ for all $x \in {\mathbb{F}}_2^n\backslash \{ \alpha ,\beta \} $; it is denoted by $s = (\alpha ,\beta )$.


\item A permutation $s$ is called {\bf APN} (Almost Perfect Nonlinear) if, for every nonzero $a \in {\mathbb{F}}_2^n$ and every $b  \in {\mathbb{F}}_2^n$, the equation
$s(x \oplus a) \oplus s(x) = b$ has at most 2 solutions.

\end{itemize}

\subsubsection{Solution}
\begin{enumerate}
  \item[{\bf Q1}]
  Let $a\in {\rm \Lambda}(g)$. Hence, $a = x\oplus y$, where $y = g(x)$ and $(x,y) = (\alpha_i, \alpha'_i)$ for some $i$. Then
  $$g(x\oplus a) = g(y) = x = y \oplus a = g(x) \oplus a.$$

  Let $a\in {\rm B}(g)$. Hence, $a = x\oplus y$, where $x,y\in {\rm FixP}(g)$. Then
  $$g(x\oplus a) = g(y) = y = x \oplus a = g(x) \oplus a.$$

  Thus, ${{\rm d}_{a ,a }}(g)\geqslant 2$ for any vector $a\in \Lambda(g) \cup {\rm B}(g)$.

  Let $g$ be an APN permutation. Then ${{\rm d}_{a ,a }}(g) = 2$. Hence, the multiplicity of all elements from $\Lambda(g)$ and ${\rm B}(g)$ is 1. Thus, $\Lambda(g) = \widehat \Lambda (g)$ and ${\rm B}(g) = \widehat {\rm B} (g)$. Note that $\Lambda(g) \cap {\rm B}(g) = \emptyset$.

  \item[{\bf Q2}]  Since $g$ is an APN permutation, then ${{\rm d}_{a ,a }}(g) \leqslant 2$. As we get in {\bf Q1}, ${{\rm d}_{a ,a }}(g) = 2$ for any vector $a\in \Lambda(g) \cup {\rm B}(g)$. Let us prove that ${{\rm d}_{a ,a }}(g) = 0$ for $a\notin \Lambda(g) \cup {\rm B}(g)$.

      Let $a$ be a nonzero vector and $x$ be a solution of $g(x  \oplus a) \oplus g(x ) = a$. Since $g$ is a permutation, then either $x\in {\rm FixP}(g)$ or $x = \alpha_i$ ($x = \alpha'_i$) for some $i$. Consider two cases:

      \begin{itemize}
      \item[{\bf 1.}] Let $x\in {\rm FixP}(g)$. Then, $g(x\oplus a) \oplus g(x) = a$ implies $g(x\oplus a) = x \oplus a$. Hence, $x\oplus a\in {\rm FixP(g)}$. As a result, $a\in B(g)$.

      \item[{\bf 2.}] Without loss of generality, let $x = \alpha_i$ for some $i$ and $y = x \oplus a$. If $y\in {\rm FixP}(g)$, then $g(x\oplus a) \oplus g(x) = a$ implies $g(x) = x$, which is a contradiction. Hence, without loss of generality, $y = \alpha'_j$ for some $j$ (so, we have $\alpha_i \oplus \alpha'_j = a$). Then
          $$g(\alpha_i \oplus a) \oplus g(\alpha_i) = a \ \Rightarrow\ g(\alpha'_j) \oplus \alpha'_i = a \ \Rightarrow\ \alpha_j \oplus \alpha'_i = a.$$
          Let us show that $\alpha'_i$ and $\alpha_j$ is also solutions. Indeed,
          $$g(\alpha'_i \oplus a) \oplus g(\alpha'_i) = g(\alpha_j) \oplus \alpha_i = \alpha'_j \oplus \alpha_i = a $$
         and
          $$g(\alpha_j \oplus a) \oplus g(\alpha_j) = g(\alpha'_i) \oplus \alpha'_j = \alpha_i \oplus \alpha'_j = a. $$
          Thus, if $i\neq j$, we get at least 3 solutions that is contradiction for the APN property of $g$. Hence, $j = i$ and $a\in \Lambda(g)$.

      \end{itemize}

  \item[{\bf Q3}] Let us prove that  $\left| {{\rm FixP}(g)} \right| \leqslant 1 + (2^{n - 1} - 1)^{1/2}$.

The involution $g$ is APN. From {\bf Q1} we have
\begin{equation}\label{E:1}
{\rm B}(g) \cap \Lambda (g) = \emptyset.
\end{equation}
Let $q = \left| {{\rm FixP}(g)} \right|$. Since $g$ is an involution, we have that $q$ is even. From equality  (\ref{E:1}) and $\Lambda (g) \cup {\rm B}(g) \subseteq {\mathbb{F}}_2^n\backslash \{ {\bf 0}\} $ it follows that
\begin{equation}\label{E:2}
\left| {\Lambda (g)} \right| + \left| {{\rm B}{\rm B}(g)} \right| \leqslant {2^n} - 1.
\end{equation}
Since $\left| {{\rm B}(g)} \right| = \binom{q}{2}$, $\left| {\Lambda (g)} \right| = {2^{n - 1}} - q/2$, we have
$$\left| {\Lambda (g)} \right| + \left| {{\rm B}{\rm B}(g)} \right| = q(q - 1)/2 + {2^{n - 1}} - q/2.$$

From inequality (\ref{E:2}),  we get
$$q(q - 1)/2 + {2^n} - q \leqslant {2^n} - 1.$$
Thus,
$$q(q - 2)/2 \leqslant {2^{n - 1}} - 1,$$
i.\,e.
$$q \leqslant 1 + {({2^{n - 1}} - 1)^{1/2}}.$$

\item[{\bf Q4}]

\begin{itemize}
\item[(a)] It could be computationally verified that $M_2 = \emptyset$ and $|M_3| = 224$.
Then, it is known~\cite{08-BrLean} that there are no APN permutations for $n = 4$. Hence, $M_4 =\emptyset$.

\item[(b)] Let us recall several definitions. A function $A:\mathbb{F}_2^n\to\mathbb{F}_2^n$ is {\it affine} if $A(x\oplus y) = A(x) \oplus A(y) \oplus A({\bf 0})$ for any $x,y\in\mathbb{F}_2^n$. Two functions $F,G:\mathbb{F}_2^n\to\mathbb{F}_2^n$ are called {\it affine equivalent} if there exist affine permutations $A_1, A_2$ such that $F = A_1\circ F\circ A_2$.
    It is easy to see that the APN permutation property of a function is an invariant under the affine equivalence.
        There exist~\cite{08-BrLean} only five the affine equivalence classes of APN permutations. Moreover, by \cite[theorem~3]{08-BrLean} only one class contains functions together with their inverses. Hence, only this class of APN permutations can contain involutions. The representative of this class is the famous inverse function over the finite field: $F(x) = x^{-1}$ for nonzero $x$ and $F(0) = 0$ (here, functions   from $\mathbb{F}_2^n$ to $\mathbb{F}_2^n$ are considered as functions over the finite field of order $2^n$).
    The inverse function is an involution. Thus, all APN involutions for $n=5$ are affine equivalent to the inverse function.

\item[(c)] There were no interesting suggestions by the participants for these open problems.

\end{itemize}

\end{enumerate}

The unique full correct solution in the first round was proposed by Henning Seidler (Germany, TU Berlin). In the second round, the best solution for 11 scores was proposed by the team of Kristina Geut, Sergey Titov, and Dmitry Ananichev (Russia, Ural State University of Railway Transport, Ural Federal University).

\hypertarget{pr-sharing}{}
\subsection{Problem ``Sharing''}

\subsubsection{Formulation}

Bob is interested in studying mathematical countermeasures to side-channel attacks on block ciphers. He found out that techniques such as special sharings of functions can be applied. Now he is thinking about the following mathematical problem in this approach.

Let ${\cal F}$ denote the set of {\bf invertible functions} ({\bf permutations}) from $\mathbb{F}_2^4$ to $\mathbb{F}_2^4$
and ${\cal F}^n$ denote the set of invertible functions from $(\mathbb{F}_2^4)^n$ to $(\mathbb{F}_2^4)^n$. Let $F \in {\cal F}^n$ be
\begin{align*}
F(x_1, x_2, \dots, x_n) = (F_1(x_1, x_2, \dots, x_n), F_2(x_1, x_2, \dots, x_n)
\dots, F_n (x_1, x_2, \dots, x_n)),
\end{align*}
with component functions $F_i:(\mathbb{F}_2^4)^n \to \mathbb{F}_2^4$, $i=1,\dots,n$.

\bigskip

For any $f \in {\cal F}$, a function $F \in {\cal F}^n$ is called a {\bf sharing} of $f$ if
$$
\sum_{i=1}^n F_i(x_1, x_2, \dots, x_n)  =
f\left(\sum_{i=1}^n x_i\right) \text{~~for all~~} (x_1, x_2, \dots, x_n) \in (\mathbb{F}_2^4)^n.
$$
Moreover, $F$ is a {\bf non-complete} sharing of $f$ if $F$ is a sharing of $f$ with the additional property that each
component function $F_i$ is independent of $x_i$.

\bigskip
Bob needs your help to study functions for which non-complete sharing exists. Find answers to the following questions!
\begin{itemize}
\item[{\bf Q1}] Let ${\cal A}$ denote the set of {\bf affine functions} from $\mathbb{F}_2^4$ to $\mathbb{F}_2^4$.
Two functions $f, g \in {\cal F}$ are {\bf affine equivalent} if there exist
$ a, b \in {\cal A}$ such that $g = b \circ f \circ a$.

Let $f, g$ be two functions in the same affine equivalence class of $\cal F$ and let $F$ be a non-complete sharing
of $f$. Derive from $F$ a non-complete sharing for $g$.
\end{itemize}

All functions of the same affine equivalence class have the same degree. It is known \cite{07-Cannier} that this equivalence relation partitions ${\cal F}$ into 302 classes: 1 class corresponds to ${\cal A}$, 6 classes contain quadratic functions, 295 classes contain cubic functions.

Also, Bob knows that when $n \geqslant 5$, there exists a non-complete sharing  for each $f \in {\cal F}$ (it can be shown by construction).
When $n=2$ a non-complete sharing exists only for the functions in~$\cal A$.
When $n=3$, non-complete sharings exist for $\cal A$ and also for 5 out of the 6 equivalence classes containing quadratic functions.
When $n=4$, non-complete sharings exist for $\cal A$, for all 6 quadratic equivalence classes and for 5 cubic classes.
\begin{itemize}
\item[{\bf Q2}] {\bf Bonus problem (extra scores, a special prize!)}

 Find a concise mathematical property that a function $f \in {\cal F}$ must have in order that
a non-complete sharing $F$ exists for $n=3,4$.

\item[{\bf Q3}] {\bf Bonus problem (extra scores, a special prize!)}

Generalize to functions over $\mathbb{F}_2^5$, $\mathbb{F}_2^6$.
\end{itemize}

\subsubsection{Solution}
\begin{itemize}
\item[{\bf Q1}]
Let $f, g$ be two functions in the same affine equivalence class of $\mathcal{F}$, that is $g=b\circ f\circ a$ for some $a,b\in\mathcal{A}$, and let $F\in\mathcal{F}^n$ be a non-complete sharing of $f$. At first, one can notice that since $f,g$ are invertible, the mappings $a,b$ must be invertible as well. Let us denote
\begin{equation*}
a(x)=Ax+a',\ \ x\in\mathbb{F}_2^4,
\end{equation*}
\begin{equation*}
b(x)=Bx+b',\ \ x\in\mathbb{F}_2^4,
\end{equation*}
where $A,B$ are nonsingular binary matrices of order $4\times4$ and $a',b'\in\mathbb{F}_2^4$.

Using components functions $\left\{F_i\right\}_{i=1}^n$ of $F$, we define the invertible function $G\in\mathcal{F}^n$ with components functions
\begin{equation*}
G_j\left(x_1,x_2,...,x_n\right)=
\begin{cases}
BF_1\left(Ax_1+a',Ax_2,...,Ax_n\right)+b',\ \ j=1,\\
BF_j\left(Ax_1+a',Ax_2,...,Ax_n\right),\ \ j\ne1,\\
\end{cases}
\end{equation*}
where $j=1,2,...,n$.

Then for any $\left(x_1,x_2,...,x_n\right)\in\left(\mathbb{F}_2^4\right)^n$, it holds
\begin{multline*}
\sum\limits_{j=1}^n{G_j\left(x_1,x_2,...,x_n\right)}=BF_1\left(Ax_1+a',Ax_2,...,Ax_n\right)+b'+\\
+\sum\limits_{j=2}^n{BF_j\left(Ax_1+a',Ax_2,...,Ax_n\right)}=B\left(\sum\limits_{j=1}^n{F_j\left(Ax_1+a',Ax_2,...,Ax_n\right)}\right)+b'=\\
=Bf\left(Ax_1+a'+Ax_2+\ldots+Ax_n\right)+b'=Bf\left[A\left(\sum\limits_{i=1}^n{x_i}\right)+a'\right]+b'=\\
=b\circ f\circ a\left(\sum\limits_{i=1}^n{x_i}\right)=g\left(\sum\limits_{i=1}^n{x_i}\right).
\end{multline*}
Therefore, the function $G\in\mathcal{F}^n$ defined as
\begin{equation*}
G\left(x_1,x_2,...,x_n\right)=\left(G_1\left(x_1,x_2,...,x_n\right),G_2\left(x_1,x_2,...,x_n\right),...,G_n\left(x_1,x_2,...,x_n\right)\right),
\end{equation*}
is a sharing of $g$.

From non-completeness of $F$ it follows that $G_j$, which is in fact an affine transformation of $F_j$, does not depend on $x_j$. Hence, $G$ is a non-complete sharing of $g$.

\item[{\bf Q2-Q3}]

These open problems were not solved completely during the Olympiad. Nevertheless, one perspective solution was proposed by the team of Victoria Vlasova, Mikhail Polyakov, and Alexey Chilikov (Bauman Moscow State Technical University). They found a sufficient condition for the existence of non-complete sharing for $n=3$. Let us describe it here.

Let $\mathrm{wt}(y)$ be the Hamming weight of a binary vector $y$. For $\sigma\in\mathbb{F}_2$, we denote
\begin{equation*}
\delta_{\sigma}(y)=
\begin{cases}
y,\ \ \sigma=1,\\
{\bf 0},\ \ \sigma=0,\\
\end{cases}
\end{equation*}
where ${\bf 0}$ is a zero vector of the same dimension as $y$.

Let $V$ be a vector space over the field $K$ and assume that for the invertible function $f:V\rightarrow V$ it holds
\begin{equation}\label{hypercube_sum}
\sum\limits_{\sigma\in\mathbb{F}_2^n}(-1)^{\mathrm{wt}(\sigma)} f\left(\sum\limits_{i=1}^n\delta_{\sigma_i}\left(x_i\right)\right)=0,
\end{equation}
then there exists a non-complete sharing for $f$. Further we conider the case $n=3$.

Indeed, for any $\left(x_1,x_2,x_3\right)\in V^3$ put
\begin{equation*}
F_1\left(x_1,x_2,x_3\right)=f\left(x_2\right)-f\left(x_2+x_3\right),
\end{equation*}
\begin{equation*}
F_2\left(x_1,x_2,x_3\right)=f\left(x_3\right)-f\left(x_1+x_3\right),
\end{equation*}
\begin{equation*}
F_3\left(x_1,x_2,x_3\right)=f\left(x_1\right)-f\left(x_1+x_2\right).
\end{equation*}

It is clear that every $F_i:V^3\rightarrow V$ does not depend on $x_i$, where $i=1,2,3$. Consider the expression
\begin{multline*}
\sum\limits_{i=1}^3F_i\left(x_1,x_2,x_3\right)=f\left(x_2\right)-f\left(x_2+x_3\right)+f\left(x_3\right)-f\left(x_3+x_1\right)+f\left(x_1\right)-f\left(x_1+x_2\right)=\\
=\sum\limits_{\sigma\in\mathbb{F}_2^3}(-1)^{\mathrm{wt}(\sigma)} f\left(\sum\limits_{i=1}^3\delta_{\sigma_i}\left(x_i\right)\right)+f\left(x_1+x_2+x_3\right)-f(0)=f\left(x_1+x_2+x_3\right)-f(0).
\end{multline*}

Without loss of generality we assume that $f(0)=0$, otherwise we can consider the initial problem for the function $g(x)=f(x)-f(0)$ with $g(0)=0$ and which, by the arguments from~{\bf Q1}, has non-complete sharing if and only if $f$ does.

Finally
\begin{equation*}
\sum\limits_{i=1}^3F_i\left(x_1,x_2,x_3\right)=f\left(x_1+x_2+x_3\right),
\end{equation*}
that concludes the proof.

It was also shown by the authors that the condition $(\ref{hypercube_sum})$ is necessary for the existence of non-complete sharing of $f$ for any $n$.

Taking $V=\mathbb{F}_2^m$ with $m=4,5,6$ and $K=\mathbb{F}_2$ one can obtain a solution of {\bf Q2, Q3} for the case $n=3$.

\end{itemize}

\hypertarget{pr-factoring}{}
\subsection{Problem ``Factoring in 2019''}

\subsubsection{Formulation}

Nicole is learning about the RSA cryptosystem. She has chosen random 500-bit prime numbers $p$ and $q$, $2^{499} \leqslant p,q < 2^{500}$, and computed $n = p\cdot q$. Being a curious and creative person, she has also combined the three numbers in funny ways. Her favorite one is an integer $h$ such that
$$h \equiv 3^{2019}p^2+5^{2019}q^2 \pmod{n^2+8\cdot 2019}.$$

Unfortunately, she has lost the paper where she wrote the two prime numbers. Luckily, she remembers $n$ and $h$. Help Nicole to recover $p$ and $q$.

\begin{align*}
n ~=~& 40763613025504836845249840044831561583564626405535158138667037 \\
& 18791672670905308860844304055285019651507728831663677166092475 \\
& 16155419756121537288444995708421977847213953345126368990185271 \\
& 10259760189356588305406519080647582874212687596214191915933827 \\
& 67252094717222418132289251314647500491996323400002019 , \\
\end{align*}
\begin{align*}
h ~=~& 78307999278336577586961528110240026923828914927526911949501196 \\
& 64549497756373569985393554661132717198368717093111812566649031 \\
& 17342818449633588647098544612151278035131454234786653136500887 \\
& 08830470996542888912418213532073622903727205396807848603735835 \\
& 72653630883685906916701587362236649126895719656663293825501223 \\
& 97088799629252601249428062432254738935764304610281613264225641 \\
& 74990272864680012560095992125783832230234589257650929348364268 \\
& 48117494065463529201859600747521892957258104033195441014023432 \\
& 36581529201392185327635674923459290749241831590661903965132514 \\
& 2154451518308886658505820006667836934411881 .
\end{align*}

\subsubsection{Solution}

This problem is based on a (simplified) variation of the Coppersmith method.

Let $m = n^2 + 8\cdot 2019$. It is a composite number with unknown factors. The idea is to find an integer $a$ such that numbers
\begin{align*}
    &a_1 = a\cdot 3^{2019}\mod{m},~\text{and}~\\
    &a_2 = a\cdot 5^{2019}\mod{m}
\end{align*}
are small enough and $a_1p^2 + a_2q^2$ exceeds the modulus $m$ by a small amount and can be recovered from $a\cdot h \mod{m}$. This can be done using the Lagrange-Gauss algorithm (which is a special case and the building block of the LLL algorithm). Let $\Lambda$ be the lattice spanned by the two vectors
\begin{align*}
v_1 &= \big(1,~~ (5^{2019}\cdot(3^{2019})^{-1} \mod m)\big),\\
v_2 &= \big(0,~~ m\big).
\end{align*}
Consider an arbitrary vector $v = (a_1, a_2)$ in this lattice. It is easy to verify that $$
    a_1p^2 + a_2q^2 \equiv a_1 \cdot h \cdot (3^{2019})^{-1} \pmod{m}.
$$
The lattice reduction guarantees to find such vector $v$ with the norm
$$
\lVert v \rVert = \sqrt{a_1^2 + a_2^2} \le 2^{(d-1)/4} (\det{\Lambda})^{1/d} = \sqrt{m}/\sqrt[4]{2},
$$
where $d=2$ is the dimension of the lattice. In particular,
$$
|a_1p^2 + a_2q^2| \le n(p^2+q^2) < n(p+q)^2 < 10 n^2,
$$
where the last two inequalities follow from balancedness of the primes (i.e., $\max(p,q) \le 2\min(p,q)$).

It follows that there exists an integer $z, |z| < 10$, such that
$$
    a_1 \cdot h \cdot (3^{2019})^{-1} \mod{m} + z m = a_1p^2 + a_2q^2.
$$
As a result, we obtain an equation in $p^2$ and $q^2$. By replacing $p=n/q$, we obtain a biquadratic equation in $q$ which is easy to solve and factor $n$.






The final solution is:
\begin{align*}
p ~=~& 20190000758781541816811298104144770223468182091751945248792088 \\
& 90921501144547048007953722271285690350264116081579241189587393 \\
& 202602664199899594021414383 , \\
q ~=~& 20190000739734941945213398056820939591822657460839955948263937 \\
& 53631669289175827851666668014167119439386543289850940734885806 \\
& 826120718179729242641026893 .
\end{align*}

The best solution was proposed by Alexey Zelenetskiy, Mikhail Kudinov, and Denis Nabokov team (Russia, Bauman Moscow State Technical University).

\hypertarget{pr-twinpeaks}{}
\subsection{Problem ``{\tt TwinPeaks3}'' (online)}

\subsubsection{Formulation}

As Bob's previous cipher {\tt TwinPeaks2} (NSUCRYPTO-2018) was broken again, he finally decided to read some books on cryptography. His new cipher is now inspired by practical ciphers, while the number of rounds was reduced a bit for better performance.

Not only the best techniques were adopted by Bob, but also he decided to enhance his cipher by security through obscurity, so the round functions are now unknown. The only thing known about these functions is that they are the same for odd and even rounds.

New Bob's cipher works as follows.
A message $X$ is represented as a binary word of length 128. It is divided into four 32-bit words $a, b, c, d$ and then the following round transformation is applied 32 times:
\begin{center}
$(a,b,c,d)\leftarrow(b,c,d,a \oplus (F_i (b,c,d)))$\\
$F_i=F_1$ for odd rounds and $F_i=F_2$ for the rest.
\end{center}
Here $F_1$ and $F_2$ are secret functions accepting three 32-bit words and returning one word; and $\oplus$ is the binary bitwise XOR. The concatenation of the final $a, b, c, d$ is the resulting ciphertext $Y$ for the message~$X$.

Agent Cooper again wants to read Bob's messages. He caught the ciphertext
$$Y = \texttt{e473f19a247429ab33b66268d57dd241}$$
(the ciphertext is given in hexadecimal notation, the first byte is \texttt{e4}).

\begin{figure}[!h]
\centering
\includegraphics[width=0.5\textwidth]{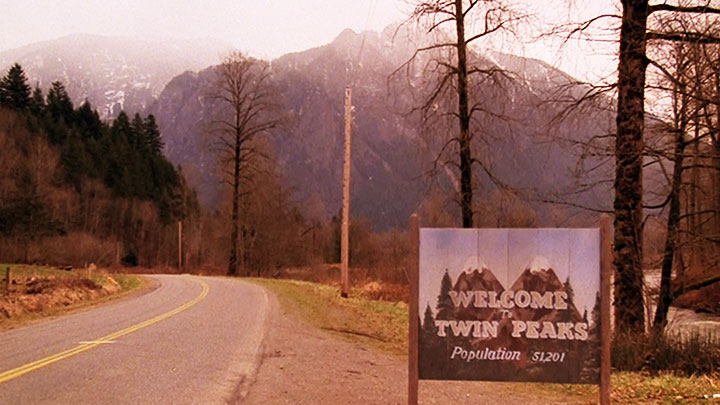}
\label{fig:twinpeaks}
\end{figure}

\newpage
He was also able to gain access to Bob's testing server with encryption and decryption routines, using the secret key. \href{https://nsucrypto.nsu.ru/archive/2019/round/2/task/4/}{\textcolor{blue}{Here}} it is \cite{twinpeaks3}. Unfortunately, the version of software available on this server is not final. So, the decryption routine is incomplete and only uses keys in the reverse order, which is not sufficient for decryption:
\begin{center}
$(a,b,c,d)\leftarrow(b,c,d,a \oplus (F_i (b,c,d)))$\\
$F_i=F_2$ for odd rounds and $F_i=F_1$ for the rest.
\end{center}
The server can also process multiple blocks of text at a time: they will be processed one-by-one and then concatenated, as in the regular ECB cipher mode of operation. Ciphertexts and plaintexts are given and processed by the server in hexadecimal notation.

Help Cooper to decrypt $Y$.

\subsubsection{Solution}

Let $f_i$ be the round transformation of round $i$:
$$ f_i:\ (a,b,c,d) \leftarrow (b,c,d,a\oplus(F_{k(i)}(b,c,d))), $$
where $k(i) = 1$ for odd $i$ and $k(i) = 2$ for the rest.

Hence, we can represent the encryption transformation $E$ as
$$E = (f_1 f_2)^{16}.$$

Let $I$ be the incomplete decryption transformation described in the problem statement.
The encryption and the incomplete decryption processes only differ in key order, so $I$ can be written in terms of $f_i$:
$$I = (f_2 f_1)^{16}.$$

The decryption transformation \(E^{-1}\) can be represented as
$$E^{-1} = (f_2^{-1} f_1^{-1})^{16},$$
where $f_i^{-1}$ is the inverse of $f_i$ and is given by the following transformation:
$$f_i^{-1}:\ (a,b,c,d) \leftarrow (d\oplus(F_{k(i)}(a,b,c)),a,b,c) $$

Thus, to apply $E^{-1}$ to the ciphertext one should be able to compute $F_1(x,y,z)$ and $F_2(x,y,z)$ that are secret. To recover these functions a \textit{slide attack} can be used.

The idea is to find words $x = (x_1,x_2,x_3,x_4)$ and $y = (y_1,y_2,y_3,y_4)$ such that $f_i(x) = y$. If such a pair is found, then $F_i$ can be found as
$$F_i(x_2,x_3,x_4) = y_4 \oplus x_1.$$

We use the following idea to find a desired pair: if $E f_i (x) = E (y)$, then $f_i (x) = y$. Let us start with $F_1$. We need a pair of $x$ and $y$ such that $E f_1 (x) = E (y)$.
This relation can be written as
\begin{center}
$(f_1 f_2)^{16} f_1 (x) = (f_1 f_2)^{16} (y)$\\
$f_1 (f_2 f_1)^{16} (x) = (f_1 f_2)^{16} (y)$\\
$f_1 I (x) = E (y)$
\end{center}

We come to a conclusion that if $f_1 I (x) = E (y)$, then $f_1 (x) = y$. The condition $f_1 I (x) = E y$ can be checked by using the definition of $f_1$: if $(I(x))_2 = (E(y))_1$, $(I(x))_3 = (E(y))_2$ and $(I(x))_4 = (E(y))_3$, then it is \textit{likely} that $f_1 I (x) = E (y)$. The probability of false positives is approximately $2^{-96}$ for random $F_i$ functions. So, it can be considered as negligible. Both $I(x)$ and $E(y)$ are available on the encryption oracle for arbitrary $x$ and $y$ as the incomplete decryption and the encryption routines respectively.

To find $F_i(a,b,c)$, let us brute force over $x$ and $y$ of the following forms: $x = (X,a,b,c)$ and $y = (a,b,c,X')$. According to the birthday paradox, a desired pair can be found in $2 * 2^{16}$ operations average (instead of $2^{32}$ if we lock $X$ or $X'$ to some constant value).

 As soon as we find such a pair $x$ and $y$, we can compute $F_1(a,b,c)$ and apply $f_1^{-1}$ to the ciphertext and decrypt the last round. Then $F_2$ can be found the same way by replacing $I$ and $E$ with each other due to the symmetry. By doing this round by round, we decrypt the whole ciphertext and get the desired message (in hexadecimal notation)
\begin{center}
{\tt acherrypieplease}
\end{center}

The reference implementation of this attack requires $2^{22}$ blocks of text to be encrypted and 10 minutes of time average. It is important to use the server's ability to process multiple blocks of text at a time to minimize the amount of HTTP requests.

Four teams successfully solved the problem using the same method.

\hypertarget{pr-curl27}{}
\subsection{{Problem ``{\tt Curl27}''}}\label{curl}

\providecommand\TT{\text{\sffamily\bfseries T}\xspace}
\providecommand{\Curl}{\text{\sffamily Curl27}\xspace}
\providecommand{\CurlF}{\text{\sffamily Curl27-f}\xspace}

\subsubsection{Formulation}

{Bob} is developing the 3OTA infrastructure and has designed a new hash function
\Curl for it. A distinguishing feature of the infrastructure is the ternary logic:
trits from the set $\TT=\{0,1,-1\}$ are used instead of bits,
ternary strings and words are used instead of binary ones. The \Curl hash
function is defined below. Its implementation in Java can be found in \cite{nsucrypto-curl}.

Find a collision for~$\Curl$, that is, different ternary strings $X$
and~$X'$ such that $\Curl(X) = \Curl(X')$.
Submit colliding strings as two lines of trits separated by commas.
An example of a (wrong!) solution is:
\begin{center}
\begin{tabular}{ll}
{\tt -1,1,0,1,1,0} \\

{\tt -1,-1,1,0,1,1,-1,0}\\
\end{tabular}
\end{center}

\noindent{\bf Description of \Curl}.
The \Curl function maps a ternary string $X$ of arbitrary length to a hash
value from~$\TT^{243}$. When hashing, an auxiliary sponge function
$\CurlF\colon\TT^{729}\to\TT^{729}$ is used. The hashing algorithm:
\begin{enumerate}
\item
Pad~$X$ with zeros to make its length a multiple of $243$.
Divide the resulting string into blocks
$X_1, X_2,\ldots, X_d\in\TT^{243}$.
\item
Prepare the state $W=W_0 W_1 W_2\in\TT^{729}$ consisting of words
$W_i\in\TT^{243}$.
Initialize the state by filling $W_0$ and $W_2$ with zeros
and~$W_1$ with the encoded initial (before padding) length of~$X$.
The length is encoded by a ternary word according to the little-endian
conventions: less significant trits go first.
For example, the length $25=1-3^1+3^3$ is presented by
the word $\underbrace{1\bar{1}01000\ldots 0}_{243}$.
Here $\bar{1}$ stands for $-1$.
\item
For $i=1,2,\ldots,d$, do: $W_0\leftarrow X_i$, $W\leftarrow\CurlF(W)$.
\item
Return $W_0$.
\end{enumerate}

\noindent{\bf Description of \CurlF}.
In \CurlF the $S$-box
$$
S\colon \TT^3 \to \TT^3,\quad
(a,b,c) \mapsto (F(a,b,c), F(b,c,a), F(c,a,b))
$$
is used. Here
\begin{align*}
F(a, b, c) &=
a^2  b^2  c + a^2 b c^2 - a b^2 c^2 + a^2 b^2 - a^2 b c + a^2 c^2 + a b^2 c\\
&- a^2 c + a b^2 - a c^2 + b^2 c + b c^2 -a^2 - b^2 + b c - c^2 - c + 1,
\end{align*}
where the calculations are carried out modulo~$3$ while the residue~$2$
is represented by the trit~$-1$.

To transform the state~$W$, $27$ rounds are performed. A round consists of 6 steps.
At each step triplets of trits of $W$ are grouped in a certain way.
Then each triplet $(a,b,c)$ is replaced with $S(a,b,c)$.

Groupings are organized as follows (see the picture below). 
At the first step, the state is divided into 3 words of 243 trits. Trits of
these words in the same positions are grouped. In the second step, the state is
divided into 9 words of 81 trits. Trits of the 1st, 2nd and 3rd words in the
same positions are grouped, then trits of the 4th, 5th and 6th words, and
so on. After that, the state is divided into words of length 27, then length 9,
then length 3 while maintaining the logic of groupings. In the last sixth step,
consecutive triplets of trits are grouped.

\bigskip

\noindent{\bf Bonus problem (extra scores, a special prize!)}.
Find a collision when the state is initialized in a different way:
now $W_0,W_2$ are not filled with zeros, the word
$\underbrace{01\bar{1}01\bar{1}\ldots01\bar{1}}_{243}$ is written in each of
them instead.

\begin{figure}[bht]
\centering
\includegraphics[width=9cm]{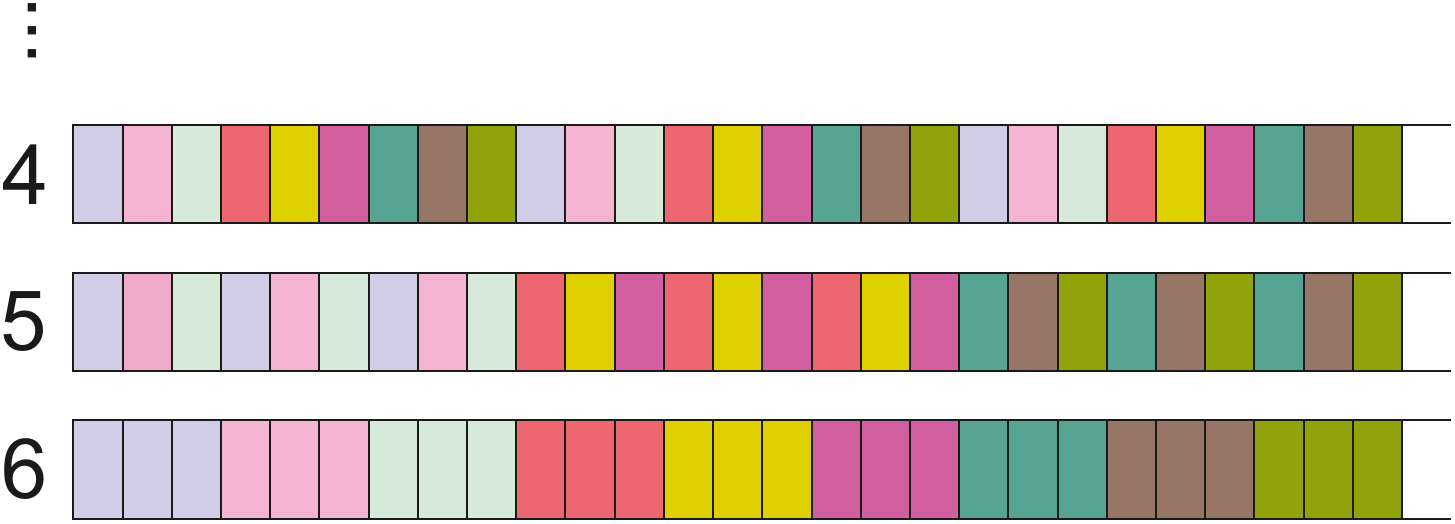}

{\small {\bf Groupings (3 last steps, grouped trits are painted the same color)}}
\label{Fig.CURL27.1}
\end{figure}

\subsubsection{Solution}

For a word~$u$ in the alphabet $\TT$, let $u^m$ be the word of $m$ copies of $u$.
Supposing $u=u_0u_1\ldots u_{n-1}$ denote $u^{[m]}=u_0^{m}u_1^{m}\ldots u_{n-1}^{m}$.
We call a word of the form $u^{[m]}$ {\it $m$-fragmented}.
\smallskip

\noindent {\bf Theorem.}
Let $m$ be a power of 3, $m\leq 729$.
The sponge function~$\CurlF$ preserves $m$-fragmentation, that is,
if $W$ is $m$-fragmented, then $\CurlF(W)$ is also $m$-fragmented.

\begin{proof}
At the $i$th step of the $\CurlF$ round function, the state~$W$ is divided
into words of length $n=3^{6-i}$, $i=1,2,\ldots,6$.
For $n\leq m$ the step function preserves equality of trits
inside fragments. It follows from the fact that $S(a,a,a)=(b,b,b)$.
For $n>m$ equality is also preserved since in each fragment trits at the
different positions are processed in the same way.
\end{proof}

Let $m$ be a small power of 3 (interesting cases are $m=3,9,27$).
Consider a ternary string $X$ of length
$$
1+3+3^2+\ldots+3^{m-1}=(3^m-1)/2.
$$
The length is given by a word of $m$ ones. Consequently, the initial state of \Curl
when processing~$X$ is $m$-fragmented (one fragment of ones, the remaining fragments of zeros).

Let us choose trits of~$X$ so as to preserve $m$-fragmentation of the state
during hashing. This is easy to do using Theorem: each full
$m$-fragment of~$X$ must have the form $\alpha^m$, $\alpha\in\TT$, and, in addition, trits of
the last (incomplete) fragment must be zero to be consistent with the padding trits.
Having achieved $m$-fragmentation of states, we automatically obtain
$m$-fragmentation of hash values. Now a hash value is determined by
$243/m$ trits, each of which is repeated $m$ times.
We can find a collision for \Curl after processing of about
$\sqrt{3^{243/m}}$ strings~$X$ of the described structure, that is,
in time of order
$$
3^m\cdot \sqrt{3^{243/m}}=3^{m+121.5/m}.
$$

The minimum of the function above is achieved at $m=9$.
During the attack with $m=9$ it is required to process approximately $\sqrt{3^{13.5}}$
strings of $9841=243\cdot 40+121$ trits each.

An example of colliding messages:
\begin{align*}
X &=0^{243\cdot 39} (101100110 101111100 101100000)^{[9]} 0^{121},\\
X'&=0^{243\cdot 39} (000011110 100111111 001000000)^{[9]} 0^{121}.
\end{align*}
This collision was found by Jeremy Jean (National Cybersecurity Agency of France), the only participant who solved
the problem.

The preservation of fragmentation is an invariant of \CurlF which allows
to decrease the dimension and thereby effectively solve the basic problem.
To solve the bonus problem, Jeremy Jean proposed to use another invariant
for \CurlF: if each part $W_0$, $W_1$, $W_2$ of the state $W$ is $3$-expanded,
then this fact also holds for $\CurlF(W)$.
Here we call a word $U\in\TT^{243}$ {\it $3$-expanded} if it has the form
$(abc)^{81}$, $abc\in\TT^3$.

In the initial state, the parts $W_0$ and $W_2$ are indeed $3$-expanded.
To comply with the invariant, the part $W_1$ representing the length of a
hashed string $X$ must have one of the forms $(ab1)^{81}$, $(a10)^{81}$ or $(100)^{81}$
(the length is nonzero and positive). As a result, $X$ consists of
at least $1+27+\ldots+27^{80}>3^{240}$ trits.

It is easy to maintain the invariant during hashing: full $243$-fragments
of $X$ must be $3$-expanded and the last incomplete fragment (if it exists)
must be filled with zeros. The resulting hash values are $3$-expanded,
there are only $27$ choices for them and a collision will surely be found after
processing only 28 strings $X$.
Of course, the attack is impractical: the time of order $3^{240}$, which is
required only for recording colliding messages, is unacceptably large even
compared to the time $3^{243/2}$ of the standard birthday attack.

\hypertarget{pr-sbox}{}
\subsection{Problem ``8-bit S-box''}

\subsubsection{Formulation}

Permutations $S$ of the set $\{0,1\}^n$ or $\mathbb{F}_2^n$ are usually called {\it $n$-bit S-boxes}. We will focus on the following cryptographic properties of S-boxes:

\begin{enumerate}[noitemsep]
\item
{\bf The (minimal) algebraic degree} of $S$, denoted by $\deg(S)$, is the minimum of algebraic degrees of all  component functions of $S$.

\item
{\bf The nonlinearity of $S$}, denoted by $\text{nl}(S)$, is the minimal Hamming distance between all component functions of $S$ and the set of all affine functions.

\item
{\bf The differential uniformity} of $S$, denoted by~$\text{du}(S)$ is the maximal number of solutions of the equation $S(x)\oplus S(x\oplus \alpha)=\beta$ for any nonzero vector $\alpha$ and any vector $\beta$.
\item
{\bf The (graph) algebraic immunity} of $S$, denoted by~$\text{ai}(S)$, is the minimal algebraic degree of all nonzero Boolean functions $f$ in $2n$ variables such that $f(x,y)=0$ for any $x\in\mathbb{F}_2^n$ and $y=S(x)$.
\end{enumerate}

In modern symmetric cryptography, S-boxes of dimension $n = 8$ are probably the most popular. For example, such an S-box is used in the AES block cipher. The characteristics of $S_{\text{AES}}$:
$$
(\deg, \text{nl}, \text{du}, \text{ai})(S_{\text{AES}})=(7,112,4,2).
$$

The value $\text{ai}(S_{\text{AES}})=2$ means that $S_{\text{AES}}$ (and the whole AES) can be compactly described by quadratic equations. This can be a weakness in the context of algebraic attacks.

Imposing the restrictions $(\deg, \text{ai})(S)=(7,3)$ (optimal values), we need to maximize $\text{nl}(S)$ and minimize $\text{du}(S)$. The current best result \cite{19-Cruz, 19-Fomin} is
$$
(\deg, \text{nl}, \text{du}, \text{ai})(S)=(7,108,6,3).
$$

{\bf Problem for a special prize!} You need to improve this result: find 8-bit S with $\text{nl}(S)>108$ and/or $\text{du}(S)<6$ while preserving $\deg(S) = 7$ and $\text{ai}(S) = 3$.

\bigskip

\noindent{\bf Remarks.} Let us recall relevant definitions.
\begin{enumerate}[noitemsep]
\item A Boolean function $f:\mathbb{F}_2^n\to \mathbb{F}_2$ can be uniquely represented in the {\it algebraic normal form}  (ANF) in the following way: $f(x) = \bigoplus_{I\in \mathcal{P}(N)} a_{I} \big(\prod_{i\in I} x_i \big),$
where $\mathcal{P}(N)$ is the power set of $N=\{1,\ldots,n\}$ and $a_I\in\mathbb{F}_2$.
\item The {\it algebraic degree} of $F$ is degree of its ANF: ${\deg}(F) = \max\{|I|:\ a_I\neq0,\ I\in\mathcal{P}(N)\}$.
\item Boolean functions of the algebraic degree not more than 1 are called {\it affine}.
\item The Hamming distance between Boolean functions $f$ and $g$ is the number of vectors $x\in\mathbb{F}_2^n$ such that $f(x)\neq g(x)$.
\item A function $S:\mathbb{F}_2^n\to\mathbb{F}_2^n$ can be given as $S = (s_1,\ldots,s_n)$, where $s_i$ is a Boolean function; a nontrivial linear combination of $s_1,\ldots,s_n$ is a {\it component} function of $S$.
\end{enumerate}

\subsubsection{Solution}

There were no valuable ideas from the Olympiad participants. The problem remains unsolved for the considered configuration of cryptographic properties. There exist several dozen of constructions, based on well-known butterfly structure, that provide current record $(7,108,6,3)$, see~\cite{19-Cruz,19-Fomin}. This leads to the idea that if candidates for improvement exist, then they are likely outside the known structures and constructions of cryptographic permutations.

\hypertarget{pr-conjecture}{}
\subsection{Problem ``Conjecture''}

\subsubsection{Formulation}

Let $\mathbb{F}_2$ be the finite field with two elements and $n$ be any positive integer
larger than or equal to~3. Let $f(X)$ be an irreducible polynomial of degree
$n$ over $\mathbb{F}_2$. It is known that the set of the equivalence classes $\beta$ of
polynomials over $\mathbb{F}_2$ modulo $f(X)$ is a finite field of order $2^n$, that
we shall denote by $\mathbb{F}_{2^n}$. It is known that different choices of the
irreducible polynomial give automorphic finite fields and such choice has
then no incidence on the algebraic problems on the corresponding fields.

{\bf Problem for a special prize!} Prove or disprove the following

\bigskip
\noindent {\bf Conjecture}.
Let $k$ be co-prime with $n$. For every $\beta \in \mathbb{F}_{2^n}$, let
$F(\beta)=\beta^{4^k-2^k+1}$. Let $\Delta=\{F(\beta)+F(\beta+1)+1;\ \beta \in
\mathbb{F}_{2^n}\}$. For every distinct nonzero $v_1,v_2$ in $\mathbb{F}_{2^n}$, we have $$|\{(x,y,z)\in \Delta^3;\ v_1x+v_2y+(v_1+v_2)z=0\}|=2^{2n-3}.$$

\bigskip

\noindent {\bf Example for $n=3$}: we can take $f(X)=X^3+X+1$, then each element $\beta$ of
the field $\mathbb{F}_{2^3}$ can be written as a polynomial of degree at most 2:
$a_0+a_1X+a_2X^2$, with $a_0,a_1,a_2\in \mathbb{F}_2$. The element 0 corresponds to
the null polynomial; and the unity, denoted by 1, corresponds to the constant
polynomial 1. We can calculate the table of multiplication in $\mathbb{F}_{2^3}$ (the
table of addition just corresponds to adding polynomials of degree at most
2); this allows calculating any power of any element of the field and check
the property.

\subsubsection{Solution}
This mathematical problem is open and difficult.
It was presented in~\cite{18-Carlet} for the first time and discussed in~\cite{18-Carlet-wcc}.
The conjecture was verified for small $n$ (odd values $n\leqslant11$, even values~$n\leqslant8$). The Olympiad participants suggested several ideas. Unfortunately, none of them gave significant advances to prove a conjecture or search for a counterexample. The team of Kristina Geut, Sergey Titov, and Dmitry Ananichev (Ural State University of Railway Transport) and the team of Alexey Zelenetskiy, Mikhail Kudinov, and Denis Nabokov (Bauman Moscow State Technical University) proved the conjecture for a particular case $k=1$. Nevertheless, this case is peculiar since the function is then quadratic and the result is known for quadratic functions. The proofs cannot be generalized to the common case.

\section{Winners of the Olympiad}\label{winners}

\noindent Here we list information about the winners of NSUCRYPTO'2019 in Tables\;\ref{1-sc},\,\ref{1-st},\,\ref{1-pr},\,\ref{2-st},\,\ref{2-pr}.

\renewcommand{\topfraction}{0}
\renewcommand{\textfraction}{0}

\begin{table}[!h]
\centering\footnotesize
\caption{{\bf Winners of the first round in school section A (``School Student'')}}
\label{1-sc}
\renewcommand{\arraystretch}{1.2}
\renewcommand{\tabcolsep}{1.6mm}
\medskip
\begin{tabular}{|c|l|p{3.0cm}|p{6.7cm}|c|}
  \hline
  Place & Name & Country, City & School &  Scores \\
  \hline
  \hline
  1 & Borislav Kirilov & Bulgaria,	Sofia &  The First Private Mathematical Gymnasium & 16 \\
   \hline
  1 & Alexey Lvov & Russia, Novosibirsk   & Gymnasium 6 & 16 \\
  \hline
  2 & Lenart Bucar	 &  Slovenia, Ljubljana	& Gymnasium Bezigrad & 15 \\
  \hline
  3&  Varvara Lebedinskaya &  Russia, Novosibirsk & The Specialized Educational Scientific Center of Novosibirsk State University &     14 \\
  \hline
  3 &  Gabriel Ericson	& Sweden, \"{O}rebro &	Tullangsskolan &   14 \\
  \hline
  Diploma   &  Vlad Coneschi &	Romania, Slatina &	Radu Greceanu National College &	11 \\
  \hline
  Diploma   &   Wang Duanyu	& Singapore, Singapore	& New Town Primary School  & 9 \\
  \hline
  Diploma   &  Vlad Ratnikov &	Russia, Yaroslavl	& School 33 of Yaroslavl &	9 \\
  \hline
  Diploma   &  Nikita Kukin	& Russia, Moscow &	Gymnasium 1540 of Moscow & 8 \\
  \hline
  Diploma   &  Michail Kostochka	& Russia, Novosibirsk &	Lyceum 130 & 8 \\
  \hline
\end{tabular}
\end{table}

\vspace{-0.5cm}

\begin{table}[!h]
\centering\footnotesize
\caption{{\bf Winners of the first round, section B (in the category ``University Student'')}}
\label{1-st}
\renewcommand{\arraystretch}{1.2}
\renewcommand{\tabcolsep}{0.8mm}
\medskip
\begin{tabular}{|c|l|l|l|c|}
  \hline
  Place & Name  & Country, City & University & Scores \\
  \hline
  \hline
  1 &   Maxim Plushkin	& Russia,	Moscow &	Lomonosov Moscow State University    & 22 \\
  \hline
  1 & Mikhail Kudinov &	Russia, Moscow & Bauman Moscow State Technical University &  21 \\
  \hline
  2 & Narendra Patel & India, Roorkee	& Indian Institute of Technology Roorkee & 19\\
  \hline
  2 & Vladimir Schavelev	& Russia, Saint Petersburg &	Saint Petersburg State University &  19\\
  \hline
  3 &Thanh Nguyen Van &	Vietnam, Ho Chi Minh City & Ho Chi Minh City University of Technology & 16\\
  \hline
  3  & Daria Grebenchuk &	Russia, Yaroslavl &	Yaroslavl State University	 & 16 \\
  \hline
  3 & Roman Gibadulin	& Russia, Yaroslavl &	Yaroslavl State University	 & 16\\
  \hline
  3 & Tuong Nguyen	& Vietnam, 	Ho Chi Minh City &  Ho Chi Minh City University of Technology & 15\\
  \hline
  Diploma & Denis Nabokov	& Russia, Moscow  & Bauman Moscow State Technical University & 14\\
  \hline
  Diploma   & Filip Dashtevski	& Macedonia, Kumanovo	& TU Delft &    14 \\
  \hline
  Diploma & Sayooj Samuel	& India, Kollam &	Amrita University  & 14\\
  \hline
  Diploma   & Paul Cotan &	Romania, Ia\c{s}i & Alexandru Ioan Cuza University & 13\\
  \hline
  Diploma & Karina Kruglik &	Belarus,	Minsk &	Belarusian State University & 13\\
  \hline
  Diploma   & Hosein Hadipour	& Iran,	Tehran	& University of Tehran	 &    13\\
  \hline
  Diploma & Polina Raspopova &	Russia, Yekaterinburg	 &	 Ural State University of Railway Transport  & 12\\
  \hline
  Diploma   & Gorazd Dimitrov	& Macedonia, Skopje	& Ecole Polytechnique  & 12\\
  \hline
  Diploma   & Diana Bespechnaya &	Russia, Moscow	& Bauman Moscow State Technical University  & 12\\
  \hline
  Diploma   & Nikolay Prudkovskiy	& Russia, Moscow	& Bauman Moscow State Technical University & 12\\
  \hline
  Diploma   & Riccardo Zanotto	& Italy, Pisa & University of Pisa  & 12\\
  \hline
  Diploma   & Dmitry Zakharov	& Russia, Moscow & National Research Nuclear University MEPhI  & 12\\
  \hline
\end{tabular}
\end{table}

\vspace{-0.5cm}

\begin{table}[!h]
\centering\footnotesize
\caption{{\bf Winners of the first round, section B (in the category
``Professional'')}}
\label{1-pr}
\medskip
\renewcommand{\arraystretch}{1.2}
\renewcommand{\tabcolsep}{1.2mm}
\begin{tabular}{|c|l|l|p{6.5cm}|c|}
\hline
Place & Name & Country, City & Organization & Scores \\
\hline
\hline
1   &   Henning Seidler & Germany, Berlin   & TU Berlin &   26 \\
\hline
2  & Samuel Tang	& Hong Kong,	Hong Kong	& Black Bauhinia  & 20 \\
\hline
2  & Madalina Bolboceanu &	Romania, Bucharest &	Bitdefender  & 20 \\
\hline
3  & Irina Slonkina &	Russia, Moscow	& National Research Nuclear University MEPhI  & 16 \\
\hline
Diploma & Harry Lee	& Hong Kong, Hong Kong	& Blocksquare Limited &  14\\
\hline
Diploma & Alexey Chilikov	& Russia,	Moscow	  & Bauman Moscow State Technical University &  14\\
\hline
Diploma & Victoria Vlasova	& Russia,	Moscow	  & Bauman Moscow State Technical University &  14\\
\hline
Diploma & Darko Ninkovic	& Serbia, Belgrade &	University of Belgrade &  13\\
\hline
Diploma & Dheeraj M Pai	& India, Chennai	& Hyperweb Media Private Limited &  13\\
\hline
Diploma & Dmitry Ananichev	& Russia, Yekaterinburg &	Ural Federal University &  13\\
\hline
Diploma &   Ekaterina Kulikova &	Germany, Munich &  & 13\\
\hline
Diploma &   George Teseleanu	& Romania,	Bucharest & Institute of Mathematics of the Romanian Academy & 12\\
\hline
\end{tabular}
\end{table}

\vspace{-0.5cm}

\begin{table}[!h]
\centering\footnotesize
\caption{{\bf Winners of the second round (in the category
``University student'')}}
\label{2-st}
\renewcommand{\arraystretch}{1.2}
\renewcommand{\tabcolsep}{0.8mm}
\medskip
\begin{tabular}{|c|p{5.0cm}|p{3.2cm}|p{5.9cm}|c|}
\hline
Place & Name  & Country, City & University &  Scores \\
\hline
\hline
1 & Alexey Zelenetskiy, Mikhail \newline Kudinov, Denis Nabokov &	Russia, Moscow	& Bauman Moscow State Technical \newline University  &       51 \\
\hline
2 & Ngoc Ky Nguyen, Dung Truong, \newline Phuoc Nguyen Ho Minh &	Vietnam,	Ho Chi Minh \newline City; France,	 Paris & Ho Chi Minh City University of \newline Technology, Ecole Normale Superieure &   43 \\
\hline
2& Thanh Nguyen Van, Quoc Bao \newline Nguyen, Ngan Nguyen	& Vietnam,\newline	Ho Chi Minh City  & Ho Chi Minh City University of \newline Technology &      40 \\
\hline
3 & Maxim Plushkin	& Russia,	Moscow &	Lomonosov Moscow State University  &       34 \\
\hline
3   & Ilya Trusevich, Maxim Bibik, \newline Alexander Shulga &	Belarus,	Minsk	&  Belarusian State University & 38 \\
\hline
Diploma & Paul Cotan,  \newline Evgnosia-Alexandra Kelesidis	& Romania, Ia\c{s}i & Alexandru Ioan Cuza University &  26\\
\hline
Diploma & Roman Sychev, Diana \newline Bespechnaya, Nikolay Prudkovskiy &	Russia, Moscow  & Bauman Moscow State Technical \newline University & 24\\
\hline
Diploma   & Vladimir Paprotski, Dmitry \newline Zarembo, Karina Kruglik &	Belarus,	Minsk	&  Belarusian State University & 21 \\
\hline
Diploma   & Vitaliy Cherkashin, Zoya \newline Tabikhanova, Evgenia Bykova	& Russia, Novosibirsk & Novosibirsk State Pedagogical University & 18 \\
\hline
\end{tabular}
\end{table}

\vspace{-0.5cm}

\begin{table}[!h]
\centering\footnotesize
\caption{{\bf Winners of the second round (in the category ``Professional'')}}
\label{2-pr}
\renewcommand{\arraystretch}{1.2}
\renewcommand{\tabcolsep}{1.0mm}
\medskip
\begin{tabular}{|c|p{4.7cm}|p{2.8cm}|p{6.4cm}|c|}
\hline
Place & Names & Country, City & Organization  & Scores \\
\hline
\hline
1   &   Irina Slonkina, Mikhail Sorokin, \newline Vladimir Bobrov &	Russia, Moscow & Bauman Moscow State Technical\newline University & 48\\
\hline
1& Kristina Geut, Sergey Titov, \newline Dmitry Ananichev &	Russia,\newline Yekaterinburg & Ural State University of Railway\newline Transport, Ural Federal University  & 46\\
\hline
2   &   Henning Seidler, Katja Stumpp & Germany, Berlin & Berlin Technical University &  42 \\
\hline
3   & Victoria Vlasova, Mikhail \newline Polyakov, Alexey Chilikov	& Russia,	Moscow	  & Bauman Moscow State Technical \newline University & 37\\
\hline
3   & Duc Tri Nguyen, Quan Doan, \newline Tuong Nguyen	& Vietnam,\newline	Ho Chi Minh City & Cryptographic Engineering Research Group, pwnphofun, Ho Chi Minh City University of Technology & 36\\
\hline
3   & Madalina Bolboceanu, \newline Andrei Mogage, Radu Titiu	& Romania, Bucharest &	Bitdefender, Alexandru Ioan Cuza \newline University & 34\\
\hline
Diploma &   Elena Kirshanova, Semyon \newline Novoselov, Nikita Kolesnikov	& Russia, Kaliningrad	 & Immanuel Kant Baltic Federal University & 28\\
\hline
Diploma &   Vyacheslav Salmanov, Evgeniya \newline Ishchukova, Nikita Kutovoy &	Russia, Taganrog & Southern Federal University & 22\\
\hline
Diploma & Jeremy Jean &	France, Paris	& National Cybersecurity Agency of France & 20\\
\hline
Diploma &  Khai Hanh Tang, Pham Phuong, \newline Yi Tu &	Singapore, \newline	Singapore &  Nanyang Technological University &  21\\
\hline
Diploma &   Harry Lee, Samuel Tang  & Hong Kong, \newline Hong Kong & Black Bauhinia    & 20\\
\hline

Diploma &  Danh Nam Tran, Thu Hien Chu  \newline Thi, Phu Nghia Nguyen &	Vietnam, \newline	Ho Chi Minh City & Ho Chi Minh City Pedagogical University, Japan Advanced Institute of Science and Technology, Ho Chi Minh City University of Technology & 20\\
\hline

\end{tabular}
\end{table}

\FloatBarrier

\newpage
\restoregeometry

\end{document}